\documentclass[aps,prb,10pt,english]{revtex4-1}
\usepackage[T1]{fontenc}
\usepackage{libertineRoman}
\usepackage{biolinum}
\usepackage{libertineMono}
\usepackage[utf8]{inputenc}
\usepackage[letterpaper]{geometry}
\usepackage{xcolor}
\usepackage{babel}
\usepackage{mathtools}
\usepackage{amsmath}
\usepackage{amssymb}
\usepackage{graphicx}
\usepackage{esint}
\PassOptionsToPackage{normalem}{ulem}
\usepackage{ulem}
\usepackage[unicode=true,
 bookmarks=false,
 breaklinks=false,pdfborder={0 0 1},backref=false,colorlinks=true]
 {hyperref}
\hypersetup{
 final,linkcolor=blue,citecolor=blue,filecolor=magenta,urlcolor=blue}

\makeatletter

\usepackage{babel}
\usepackage{setspace}

\usepackage{relsize}
\usepackage{caption}
\usepackage[]{ulem}
\usepackage{tabto}
\usepackage{enumitem}
\usepackage{tikz}
\usepackage{stackengine}
\usepackage{scalerel}
\usepackage{notoccite}

\allowdisplaybreaks
\usepackage{wasysym}
\usepackage{bbm}
\usepackage{bbold}

\DeclareTextSymbolDefault{\textquotedbl}{T1}

\DeclareMathAlphabet{\mathpzc}{OT1}{pzc}{m}{it}
\DeclareFontFamily{OT1}{pzc}{}
\DeclareFontShape{OT1}{pzc}{m}{it}{<-> s * [1.05] pzcmi7t}{}
\DeclareMathAlphabet{\mathpzc}{OT1}{pzc}{m}{it}

\newcommand{\rightGF}{{
    \tikz[line cap=round,x=1ex,y=1ex,line width=0.3pt]
    {\draw (.9,0) |- (0,1.35) (0.55,0) ;}}}

\newcommand{\leftGF}{{
    \tikz[line cap=round,x=1ex,y=1ex,line width=0.3pt]
    {\draw (.9,0) |- (1.8,1.35) (0.55,0) ;}}}

\newcommand{\Tr}[0]{\textnormal{Tr}}
\newcommand{\hata}[0]{\hat{a}}

\newcommand{\hatc}[0]{\hat{c}}
\newcommand{\hatd}[0]{\hat{d}}
\newcommand{\hatH}[0]{\hat{H}}
\newcommand{\hatU}[0]{\hat{\mathcal{U}}}
\newcommand{\hatN}[0]{\hat{N}}
\renewcommand{\'}[0]{^\prime}

\newcommand{\hatT}[0]{\hat{\mathcal{T}}}
\newcommand{\barU}[0]{\overline{\mathcal{U}}}
\newcommand{\rhohat}[0]{\hat{\rho}}
\newcommand{\Gbar}[0]{\overline{G}}

\@ifundefined{showcaptionsetup}{}{
    \PassOptionsToPackage{caption=false}{subfig}
}

\newsavebox{\@brx}
\newcommand{\llangle}[1][]{\savebox{\@brx}{\(\m@th{#1\langle}\)}
  \mathopen{\copy\@brx\kern-0.5\wd\@brx\usebox{\@brx}}}
\newcommand{\rrangle}[1][]{\savebox{\@brx}{\(\m@th{#1\rangle}\)}
  \mathclose{\copy\@brx\kern-0.5\wd\@brx\usebox{\@brx}}}

\makeatother

\begin{document}
\title{Describing non-Hermitian dynamics using a Generalized Three-Time NEGF
for a Partition-free Molecular Junction with Electron-Phonon Coupling}
\author{M. A. Lane$^{1}$, L. Kantorovich$^{1}$}
\affiliation{\singlespacing{}$^{1}$Department of Physics, King's College London,
	Strand London, WC2R 2LS, United Kingdom}

\begin{abstract}
In this paper we develop the Non-Equilibrium Green's Function (NEGF)
formalism for a dissipative molecular junction that consists of a
central molecular system with one-dimensional electronic transport
coupled to a phonon environment and attached to multiple electronic
leads. Our approach is partitionless - initial preparation of the
system places the whole system in the correct canonical equilibrium
state - and is valid for an external bias with arbitrary time dependence.
Using path integrals as an intermediary tool, we apply a two-time
Hubbard-Stratonovich transformation to the phonon influence functional
with mixed real and imaginary times to obtain an exact expression
for the electronic density matrix at the expense of introducing coloured
Gaussian noises whose properties are rigorously derived from the environment
action. This results in a unique stochastic Hamiltonian on each branch
of the Konstantinov-Perel' contour (upper, lower, vertical) such that
the time evolution operators in the Liouville equation no longer form
a Hermitian conjugate pair, thus corresponding to non-Hermitian dynamics.
To account for this we develop a generalized three-time NEGF which
is sensitive to all branches of the contour, and relate it to the
standard NEGF in the absence of phonons via a perturbative expansion
of the noises. This approach is exact and fully general, describing
the non-equilibrium driven dynamics from an initial thermal state
while subject to inelastic scattering, and can be applied to non-Hermitian
dynamics in general.
\end{abstract}

\maketitle

\section{Introduction}

In the age of nanofabrication and molecular devices, the electronic
transport properties of molecular structures and one-dimensional materials
has become an appealing theoretical question, rooted in its applications
to electronic engineering\cite{joachim1997atomic}. Of particular
interest are conducting structures where the strong confinement of
electrons across two dimensions is of the order of atomic diameters,
rendering current flow to be effectively one-dimensional; henceforth,
such structures shall be referred to as \textit{molecular junctions}.
The conductance of various current-carrying molecular junctions has
been measured using scanning probe spectroscopy or nanolithography,
with measurements including: scanning probe spectroscopy for single
molecules absorbed onto surfaces\cite{joachim1995electronic,joachim1997electromechanical,park2000nanomechanical,PhysRevLett.87.126801,neel2007controlled},
organic molecules bonded to electrodes\cite{reed1997conductance}
or embedded into self-assembling mono-layers\cite{bumm1999electron},
controlled absorption onto silicon\cite{wolkow1999controlled}, carbon
nano-tubes\cite{frank1998carbon,tans1998room,tans2000potential,venema1999imaging,liang2002shell},
macroscopic electrodes obtained via nanolithography\cite{frank1998carbon,zhou2000intrinsic,porath2000direct,ginger2004evolution,martin2008lithographic,pimpin2012review},
mechanically controllable break junctions\cite{reed1997conductance,kergueris1999experimental,he2006measuring,venkataraman2006single,martin2008lithographic,wang2019advance},
and simultaneous measurement of charge and heat transport through
single molecules\cite{widawsky2012simultaneous}. Determining the
current in such molecular junctions is not a purely electronic problem:
the electrons interact inelastically with vibrations of the atomic
lattice so that the properties of the junction must be understood
in the context of open quantum systems, where dissipative effects
play an important role in the system's properties, for example, in
scanning tunneling microscopy where an atomic chain forms at the contact
between the tip and sample with vibrations\cite{zhitenev2002conductance}.

The advancement of theoretical approaches for actually calculating
these electronic transport properties was kick-started by the development
of phenomenological models for elastic transport in static junctions
due to Landauer and Buttiker (LB) in the LB formalism\cite{landauer1957spatial,landauer1987electrical,buttiker1992scattering,imry1999conductance}
which relates the scattering properties of a conductor to its conductance.
More generally, considering elastic transport with rigid atoms has
led to an ontogeny of scattering approaches\cite{sautet1988electronic,joachim1996length,magoga1997conductance,buttiker2006scattering,moskalets2011scattering,mujica2000molecular,nitzan2003electron,doi:10.1063/1.3146905},
predominantly for one dimensional transport, but which are elastic
in the sense that they consider electron-electron interactions in
the absence of any dissipation/inelastic processes, with some notable
exceptions\cite{galperin2006inelastic,frederiksen2007inelastic,pastawski1991classical,segal2000activated,PhysRevB.63.125422,PhysRevLett.83.452}.
In fact, formally inelastic effects can be included to all orders
in electron-phonon coupling within the multichannel scattering method
\cite{Ness-JPCM-2006,PhysRevLett.83.452,dash2010nonequilibrium}.
Other elastic methods include quantum master equations\cite{harbola2006quantum,esposito2009transport}
and the Non-Equilibrium Green's Function (NEGF) formalism \cite{diventra_2008,2003pngf.conf,Hirsbrunner_2019,doi:10.1063/1.5145210,KadanoffBaym1962,Keldysh:1964ud,doi:10.1063/1.1664616,Langreth1976,DANIELEWICZ1984239,stefanucci_vanleeuwen_2013},
where the latter represents a powerful generalization of the scattering
matrix method\cite{ami2002intramolecular,stefanucci_vanleeuwen_2013,arrachea2006relation}.
Crucially, first principles approaches based on NEGFs such as DFT
have been very successful at describing the electronic properties
for a wide range of systems\cite{soler2002siesta,rungger2008algorithm,smidstrup2019quantumatk},
and are easily combined with elastic scattering. Of course, the reality
is that atomic lattice vibrations (phonons) and inelastic electron-phonon
interactions are fundamental to any description of a molecular junction\cite{RevModPhys.89.015003,Galperin-Rathner-Nitzan-rev-JPCM-2007}
at finite temperature since the coupling between electrons and phonons
is strongly enhanced for one-dimensional and molecular scale systems.
The NEGF formalism has proven a fruitful method for the inclusion
of interactions within the junction \cite{PhysRevB.82.085426,ness2011nonequilibrium,ness2012many,frederiksen2004inelastic,frederiksen2007inelastic,lu2012current,Paulsson-JPCS-2006,Paulsson-PRL-2008,Galperin-Rathner-Nitzan-rev-JPCM-2007},
in particular for inelastic effects. In fact, the latter effects are
naturally treated within the NEGF using diagrammatic methods \cite{dash2010nonequilibrium,dash2011nonequilibrium}.

Using the Feynman-Vernon influence functional formalism\cite{feynman2000theory},
the effect of the phonon environment on the electronic open system
can be calculated exactly using path integrals. Specifically, the
quantum-mechanical propagators are expressed as phase-weighted sums
over trajectories, where the phase associated with each trajectory
is proportional to the action of that trajectory in the classical
system\cite{feynman2010quantum}. This approach has since been greatly
expanded upon\cite{grabert1988quantum,smith1987generalized,makri1989effective,allinger1989influence}
and applied to many open quantum systems in first quantization, focusing
on the rigorous derivation of quantum Langevin equations for the reduced
density matrix\cite{caldeira1983path,sebastian1981classical,leggett1987dynamics,ford1987quantum,gardiner1988quantum}
or stochastic Liouville von-Neumann equations\cite{stockburger2004simulating,mccaul2017partition,PhysRevB.101.224306}
via the application of a Hubbard-Stratonovich (HS) transformation\cite{Stratonovich-1958,PhysRevLett.3.77}.
The obtained equations of motion for the density matrix or wavefunction\cite{PhysRevB.101.224306,matos2020efficient,PhysRevLett.88.170407,diosi1998non,moodley2009stochastic,breuer2009stochastic,orth2013nonperturbative,ruan2018unravelling}
are stochastic in the sense that they contain coloured Gaussian noises,
but whose properties are analytically derived from the propagator
path integral rather than being introduced artificially. These schemes
all exhibit non-Hermitian dynamics associated with a Liouville equation
where the time evolution operators do not form a Hermitian conjugate
pair, a feature which manifests in dynamics which does not preserve
the trace and can lead to numerical instability\cite{PhysRevB.101.224306,matos2020efficient}.
Extending this procedure to an electronic open system interacting
with the nuclear lattice and heat bath \cite{hedegrd1987light} has
only recently been done for molecular junctions and applied to thermal
and electronic transport\cite{lu2012current,PhysRevB.98.014307}.
The initial condition in this approach has the potential to be generalised
so that the electronic and phonon sub-spaces are not partitioned but
are jointly thermalized \cite{grabert1988quantum,smith1987generalized,grabert1987quantum,PhysRevB.95.125124},
though the issue of non-Hermicity remains a serious point.

In this work, we integrate out the phonon environment directly and
generalise the NEGF to account for the resulting non-Hermicity; even
with the tools of the NEGF formalism, the inclusion of phonons leading
to non-Hermitian dynamics is formidable. Therefore, our approach is
to marry together the advantages of the path integral and NEGF representations
of the system, using path integrals to integrate out the atomic vibrations
exactly, while the NEGF allows us to construct a consistent and elegant
framework for the electronic dynamics in the presence of phonons.
The result is the reduction of the system to an electron-only problem
in which phonons have been replaced with Gaussian noises in the Hamiltonian,
achieved via the application of a generalized HS transformation with
respect to two times rather than one\cite{PhysRevLett.88.170407,PhysRevB.95.125124,PhysRevB.101.224306,kantorovich2018nonadiabatic}:
one real time associated with the open system dynamics, and one imaginary
time associated with thermal preparation. This provides an exact procedure
for the joint thermal preparation of the electronic and phonon subsystems
together in the correct canonical equilibrium state, ensuring that
quantum coherence is retained with no adiabatic separation between
electronic states and phonons, even at the initial time. The appearance
of the Gaussian noises and non-Hermicity of the Hamiltonian requires
a generalisation of the NEGF and the Kadanoff-Baym equations of motion,
resulting in a stochastically unravelled\cite{Breuer_2009,PhysRevA.79.042103,doi:10.1080/00268976.2018.1456685,PhysRevB.101.224306}
three-time Green's function in which time reversibility with respect
to real times is broken \cite{Brandbbyge-1995,lu2012current,kantorovich2018nonadiabatic,PhysRevB.101.165408}.
We emphasise that this three-time NEGF is a rather different kind
of NEGF and that its calculation requires the introduction of new
self-energies, auxiliary functions, and use of the generalized Langreth
rules\cite{PhysRevB.101.165408}.

The purpose of this paper is to present a generalization of the NEGF
formalism capable of handling non-Hermitian dynamics, such as in a
molecular junction in the presence of electron-phonon coupling within
the framework of stochastic unravelling. The paper is organised as
follows. We begin in Section \ref{sec:model} with a description of
the model for a current junction that includes a phonon environment
in the central region which is coupled to electrons. The result of
applying the stochastic unravelling procedure to this model, transforming
phonons into coloured Gaussian noises, is presented in Section \ref{sec:HS},
with a detailed derivation in Appendix \ref{A:influence}. We then
develop the three-time NEGF which accounts for the additional branch
dependence introduced by stochasticity/non-Hermicity in Sec \ref{sec:branchdependence},
followed by a series expansion appropriate for numerical simulation
for the three-time NEGF in terms of the noises and phonon-free NEGF
in Section \ref{sec:W} with a summary of the Generalized Langreth
Rules in Appendix \ref{A:langreth}. An expression for the non-linear
current response to an external bias on the leads in the presence
of inelastic scattering is derived in Section \ref{sec:current},
with a subtlety of the derivation explained in Appendix \ref{A:Girsanov},
and expressions for the components of the self energies given in terms
of energy integrals in Appendix \ref{A:selfenergies}. Finally, in
Sections \ref{sec:discussion} and \ref{sec:conclusion}, we present
a discussion of the overall procedure and our conclusions, respectively.
At the time of publication, no implementation of this method is available
so calculations will not be presented here; this is left for future
work.

\section{Model}

\label{sec:model} 
\begin{figure}[ht]
\centering \includegraphics[width=0.5\linewidth]{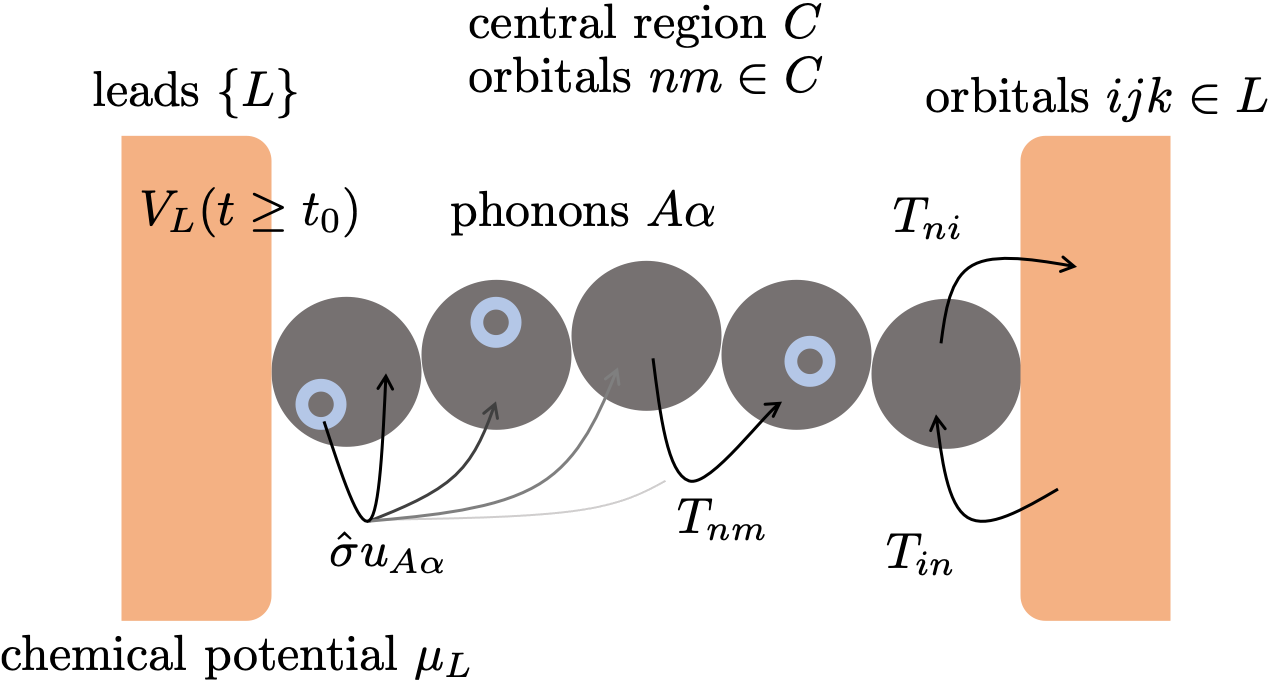} \captionsetup{width=.8\linewidth}
\caption{(Colour online) Schematic of the molecular junction with an atomic
chain (grey circles) as the central region connected to an arbitrary
number of semi-infinite leads (orange blocks, only two are shown);
the $L^{th}$ lead has chemical potential $\mu_{L}$ and external
bias $V_{L}$ which is switched on at $t_{0}$. Phonons (blue rings)
in the central region couple to electronic states in the central region
via the electronic coupling operator $\hat{\sigma}$. \label{fig:junction}}
\end{figure}

We consider a molecular junction comprised of an
interacting central region $C$ connected to any number of non-interacting
leads $\{L\}$, where each lead is under the influence of an external
time-dependent spatially homogeneous bias potential $V_{L}(t)$. This
set-up is depicted schematically in Figure \ref{fig:junction}. The
system Hamiltonian in the absence of phonons (denoted by the superscript
0) is given by 
\begin{align}
\hatH^{0}(t) & =\sum_{L,i\in L}\left[\epsilon_{i}+V_{L}(t)\right]\hatc_{i}^{\dagger}\hatc_{i}+\sum_{mn\in C}T_{mn}\hatd_{m}^{\dagger}\hatd_{n}+\sum_{L,i\in L;\,n\in C}\left(T_{ni}\hatd_{n}^{\dagger}\hatc_{i}+T_{in}\hatc_{i}^{\dagger}\hatd_{n}\right)\label{eq:H01}\\
 & =\sum_{L}\hatH_{L}(t)+\hatH_{C}^{0}+\sum_{L}\hatH_{LC},\label{eqH02}
\end{align}
where $\hatH_{L}(t)$ is the Hamiltonian of the $L^{th}$ lead which
includes the bias $V_{L}(t)$, $\hatH_{C}^{0}$ is the Hamiltonian
of the central region which refers to hopping events between eigenstates
$n$ and $m$, and $\hatH_{LC}$ contains the coupling of the $L^{th}$
lead to the central region. Here, $\hatc_{i}^{\dagger}(\hatc_{i})$
creates(annihilates) a non-interacting electron with energy eigenvalue
$\epsilon_{i}$ for any $i\in L$ in the $L^{th}$ lead, while an
interacting electron in the central region $C$ of the electronic
level $n$ is created(annihilated) by the operators $\hatd_{n}^{\dagger}(\hatd_{n})$.
For the sake of clarity, electronic state indices will be restricted
to specific subsystems within the molecular junction, so that $i,j,k\in L$,
while $n,m\in C$, and $p,q\in C\wedge\{L\}$.

In first quantization, atomic vibrations can be introduced to the
Hamiltonian Eq. \eqref{eq:H01} via each atom $A$ (of mass $m_{A}$)
displacement coordinate $u_{A\alpha}$ defined relative to the equilibrium
position within the lattice $\alpha$, and the conjugated momentum
$p_{A\alpha}$ operator. The resulting modification to the central
region Hamiltonian $\hatH_{C}^{0}$ is of the form 
\begin{align}
\hatH_{C}=\hatH_{C}^{0}+\frac{1}{2}\sum_{AA\'\alpha\alpha\'}\left[\frac{p_{A\alpha}^{2}}{2m_{A}}\delta_{AA\'}\delta_{\alpha\alpha\'}+\Lambda_{AA\'}^{\alpha\alpha\'}u_{A\alpha}u_{A\'\alpha\'}\right]+\hatH_{el-ph},\label{eq:centralHwithphonons}
\end{align}
where $\boldsymbol{\Lambda}=\left\Vert \Lambda_{AA\'}^{\alpha\alpha\'}\right\Vert $
is the force-constant matrix. The second term describes harmonic phonons
in the central region, and the third term $\hatH_{el-ph}$ describes
the electron-phonon interaction, 
\begin{gather}
\hatH_{el-ph}=-\sum_{A\alpha}\sum_{nm}\mathcal{V}_{nm}^{A\alpha}\hatd_{n}^{\dagger}\hatd_{m}u_{A\alpha}=-\sum_{A\alpha}\hat{\sigma}_{A\alpha}u_{A\alpha},\label{eq:Hel-ph}
\end{gather}
where $\mathcal{V}_{nm}^{A\alpha}=\langle\phi_{n}\vert\mathcal{V}^{A\alpha}(\textbf{r})\vert\phi_{m}\rangle$
are the matrix elements of the coupling potential $\mathcal{V}^{A\alpha}(\textbf{r})$
on the orbitals $\{\phi_{n}(\textbf{r})\}$ in the central region,
and $\hat{\sigma}_{A\alpha}=\sum_{nm}\mathcal{V}_{nm}^{A\alpha}\hatd_{n}^{\dagger}\hatd_{m}$
is the electronic coupling operator to the $A\alpha$ displacement.

At thermal equilibrium, the total density matrix of the electronic
and phonon subsystems together is given by 
\begin{align}
\hat{\rho}_{tot}(t_{0})=\frac{1}{Z_{tot,0}}e^{-\beta\hat{\mathcal{H}}^{0}},
\end{align}
where $\hat{\mathcal{H}}^{0}=\hatH^{0}(t_{0})-\mu\hatN$ is the Hamiltonian
of the Grand Canonical Ensemble at $t_{0}$, characterised by the
chemical potential $\mu$ and the number operator $\hatN$, and $Z_{tot,0}=\Tr\left[e^{-\beta\hat{\mathcal{H}}^{0}}\right]$
is the partition function of the total system.

Note that this is a phenomenological description of phonons in a coupled
electron-phonon system since the force-constant matrix is already
defined in our Hamiltonian (and thus so are the harmonic frequencies);
the full characterisation of lattice vibrations which are caused by
ion-electron interactions \cite{hedin1970effects,maksimov1975self,RevModPhys.89.015003}
goes well beyond the scope of this paper. Instead we have assumed
that when the lattice is properly dressed with electrons, lattice
vibrations which are harmonic emerge and couple with the electrons
linearly with respect to their displacements but arbitrarily with
respect to the electrons, i.e. there are no limitations applied to
the coupling strength.

It is also assumed that up until $t_{0}$ the total system was in
thermodynamic equilibrium characterized by the chemical potential
$\mu_{L}=\mu$ (the same for all leads) and inverse temperature $\beta$,
and not in a partitioned state, before each lead $L$ was subjected
to the potential $V_{L}(t)$ with subsequently arbitrary time dependence.
The physical reality \cite{verzijl2013applicability} of switching
on the bias at $t_{0}$ means that the energy levels in the $L^{th}$
lead are shifted by $V_{L}(t_{0})$, causing a rearrangement of electrons
in the junction and the leads, with electronic screening ensuring
that the internal electric field well inside any leads will be zero.
Consequently, any potential difference introduced as a result of the
bias will be confined to the central region. This is avoided by choosing
a sufficiently large central region so that the boundary layer of
each lead which is most affected by the charge redistribution is incorporated
directly into the central region and any modulation of the bias by
the charge redistribution can be neglected, with $V_{L}(t)$ once
again being uniform within the $L$-th lead.

\section{Unravelling the Phonon influence Functional}

\label{sec:HS} The central region Hamiltonian including phonons Eq.
\eqref{eq:centralHwithphonons} corresponds to harmonic bath degrees
of freedom $\{A\alpha\}$ coupled to electronic states $\{nm\}$ in
the central region, where the coupling strength is arbitrary in the
central region's coordinates but linear in the bath displacement.
Although Eq. \eqref{eq:centralHwithphonons} is presented in second
quantization, there is of course a corresponding Hamiltonian in first
quantization which consists of an electronic sub-system made up of
one-particle \textit{bra-ket} states coupled to a classical harmonic
bath. As a result, the first quantization version of Eq. \eqref{eq:centralHwithphonons}
is a more general form of the Caldeira-Leggit Hamiltonian\cite{caldeira1983path}
for which the process of stochastic unravelling is well known\cite{feynman2000theory,grabert1988quantum,mccaul2017partition}.
Applied to Eq. \eqref{eq:centralHwithphonons}, stochastic unravelling
replaces the sum over phonon degrees of freedom in the electron-phonon
interaction of Eq. \eqref{eq:Hel-ph} with a stochastic Hamiltonian
that couples the central region electron states to a set of stochastic
potential fields in the form of coloured noises, with the physical
properties of the system being recovered exactly by averaging over
all possible manifestations of these noises.

\begin{figure}[ht]
\centering \includegraphics[width=0.5\linewidth]{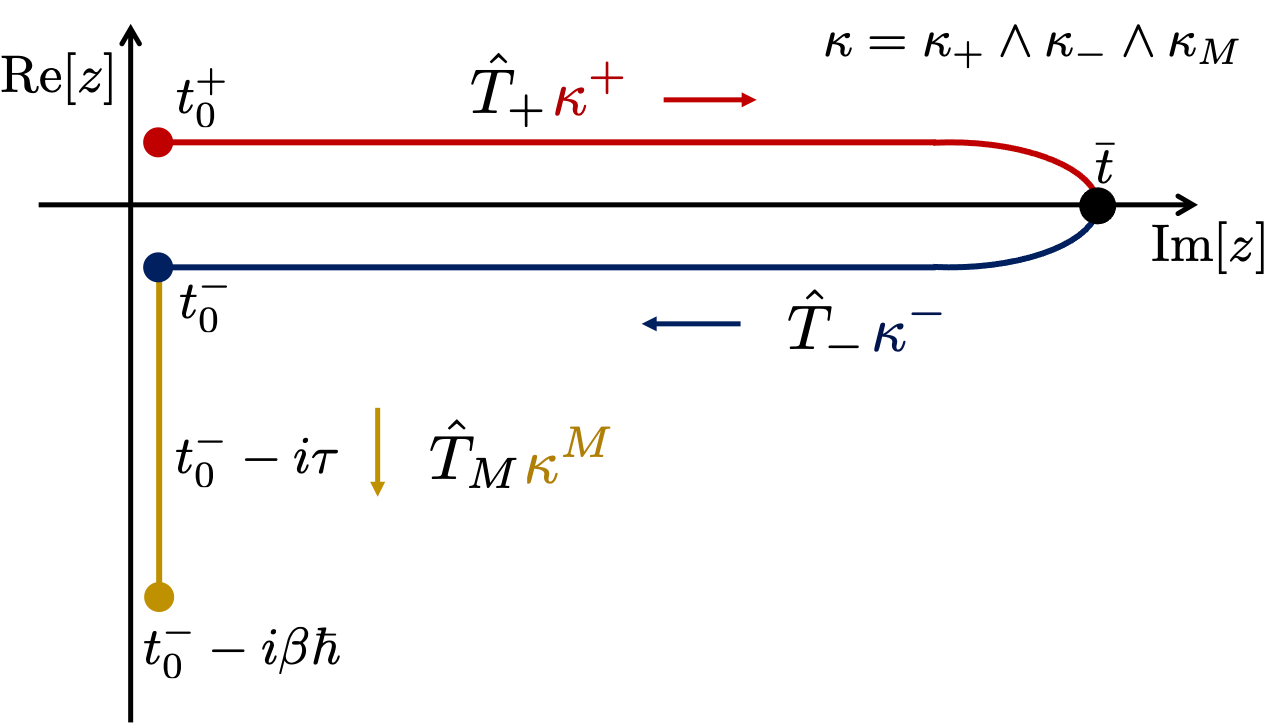} \captionsetup{width=.8\linewidth}
\caption{(Colour online) The Konstantinov-Perel' contour $\kappa$ for the
complex time $z=t-i\tau$ with $t\in(t_{0},\overline{t})$ on the
real axis and $\tau\in(0,\beta\hbar)$ on the imaginary axis, where
$\overline{t}$ is the largest (right-most) time, referred to as the
\textit{observation time}. The total contour $\kappa$ is made up
of three branches: the upper horizontal branch $\kappa^{+}$ with
chronological (ascending) real time ordering enforced by the operator
$\hatT_{+}$, the lower horizontal branch $\kappa^{-}$ with anti-chronological
(descending) real time ordering enforced by the operator $\hatT_{-}$,
and the vertical Matsubara branch $\kappa^{M}$ with chronological
time ordering enforced by $\hatT_{M}$ for the imaginary time $\tau$,
from 0 to $\beta\hbar$. \label{fig:contour}}
\end{figure}

To do this (see Appendix \ref{A:influence}), we derive an exact expression
for the reduced density matrix by taking the partial trace over the
atomic vibrations by means of the path integral method, expressing
the influence functional in the quadratic (bi-linear) form needed
to apply the HS transformation to introduce the noises. This reduced
density matrix is therefore reduced with respect to the atomic displacements,
so can be thought of as the electron-only density matrix. By then
returning to the operator language from the path integrals representation,
two stochastic propagators can be introduced, $\hatU^{\pm}$, which
enable one to write an exact expression for the time evolution of
the reduced (electronic) density matrix from its initial value, $\widetilde{\rho}_{0}$,
at time $t_{0}$ (as a formal solution of a stochastic Liouville equation)
as 
\begin{align}
\widetilde{\rho}(\overline{t})=\hatU^{+}(\overline{t},t_{0}^{+})\widetilde{\rho}_{0}\hatU^{-}(t_{0}^{-},\overline{t}),\label{eq:Liouville_noises}
\end{align}
where the tilde has been introduced to denote the fact that this is
not the physical density matrix but a stochastic one corresponding
to a single realization of the noises. Physical quantities are only
recovered after averaging over the noises and appropriate normalisation
(see below) Eq. \eqref{eq:Liouville_noises} corresponds to stochastic
dynamics along the horizontal branches of the Konstantinov-Perel'
contour $\kappa$ depicted in Figure \ref{fig:contour}, first evolving
chronologically along $\kappa^{+}$ by $\hatU^{+}$ up to the \textit{observation
time} $\overline{t}$, introduced here as the right-most real time
on the contour, before evolving anti-chronologically along $\kappa^{-}$
by $\hatU^{-}$. These propagators take the form 
\begin{equation}
\hatU^{\pm}(t,t\')=\hatT_{\pm}\exp\left\{ \mp\frac{i}{\hbar}\int_{t\'}^{t}dt_{1}\hatH^{\pm}(t_{1})\right\} ,\label{eq:U}
\end{equation}
where the Hamiltonian $\hatH^{\pm}$ is now branch dependent, 
\begin{gather}
\hatH^{\pm}(t)=\hatH^{0}+\hatH_{el-ph}^{\pm}(t)\label{eq:unravelledH}\\
\hat{H}_{el-ph}^{\pm}=-\sum_{A\alpha}\left[\eta_{A\alpha}(t)\pm\frac{\hbar}{2}\nu_{A\alpha}(t)\right]\hat{\sigma}_{A\alpha}.\label{eq:Hpm}
\end{gather}
This corresponds to removing the harmonic phonon part from Eq. \eqref{eq:centralHwithphonons}
and replacing the original electron-phonon interaction Hamiltonian
with an unravelled coupling term $\hat{H}_{el-ph}^{\pm}$ that contains
the sets of noises $\{\eta_{A\alpha}(t)\}$ and $\{\nu_{A\alpha}(t)\}$,
so that Eq. \eqref{eq:unravelledH} is for electrons only. Crucially,
the stochastic propagators on either side of $\widetilde{\rho}_{0}$
in Eq. \eqref{eq:Liouville_noises} are not each other's Hermitian
conjugate, that is, $\hatU^{+}$ is not the Hermitian conjugate of
$\hatU^{-}$. As such, the dynamics described in Eq. \eqref{eq:Liouville_noises}
is not Hermitian, with Hermicity only being recovered after taking
the stochastic average $\langle\ldots\rangle_{\xi\overline{\xi}}$.
This represents a significant deviation from standard NEGF theories,
and is the main feature of this approach which requires a generalization
of the NEGF.

Similarly, the equilibrium density matrix $\widetilde{\rho}_{0}$
at $t_{0}$ is obtained by a stochastic evolution in imaginary time,
\begin{align}
\overline{\rho}(\tau)=\barU(\tau,0),
\end{align}
with $\overline{\rho}(\beta\hbar)=\widetilde{\rho}_{0}$, where the
over-bar denotes imaginary time stochasticity, and 
\begin{equation}
\barU(\tau,\tau\')=\hatT_{M}\exp\left\{ -\frac{1}{\hbar}\int_{\tau\'}^{\tau}d\tau_{1}\mathcal{\hatH}(\tau_{1})\right\} \label{eq:Ubar}
\end{equation}
is a stochastic propagator in imaginary time from $\tau'$ and $\tau$
along the vertical branch of the contour $\kappa^{M}=\left\{ \tau\in[0,\beta\hbar]\right\} $
(see Figure \ref{fig:contour}) serving to thermalize the total system
into the correct initial canonical equilibrium state corresponding
to a particular realization of the imaginary time noises, $\{\overline{\mu}_{A\alpha}(\tau)\}$.
The Hamiltonian on the vertical branch inherits this set of imaginary
time noises, which replace the corresponding displacement operators
in the electron-phonon coupling term:
\begin{gather}
\mathcal{\hatH}(\tau)=\mathcal{\hatH}^{0}+\overline{{H}}_{el-ph}(\tau),\label{eq:H^M}\\
\overline{H}_{el-ph}(\tau)=-\sum_{A\alpha}\overline{\mu}_{A\alpha}(\tau)\hat{\sigma}_{A\alpha},\label{eq:Hbar}
\end{gather}
so that there are three sets of noises $\{\eta_{A\alpha}(t)\}$, $\{\nu_{A\alpha}(t)\}$
and $\{\overline{\mu}_{A\alpha}(\tau)\}$, with three noises $\eta_{A\alpha},\nu_{A\alpha},\overline{\mu}_{A\alpha}$
per atomic displacement $u_{A\alpha}$.

Collectively, Eqs. \eqref{eq:Liouville_noises}-\eqref{eq:Hbar} correspond
to evolution via a complex time propagator $\hatU(z,z\')$, where
$z,z\'$ can take any position on the contour, with contour ordering
operator $\hatT_{\kappa}$ which considers ascending times to run
from $t_{0}^{+}$ to $\overline{t}$, then $\overline{t}$ to $t_{0}^{-}$,
then $t_{0}^{-}$ to $t_{0}^{-}-i\beta\hbar$ so satisfies $\hatT_{\pm}$
and $\hatT_{M}$ on the horizontal and vertical branches, respectively.
It will be useful to introduce the general creation(annihilation)
operators $\hata_{p}^{\dagger}(\hata_{p})\in\left\{ \hatc_{i}^{\dagger}(\hatc_{i}),\hatd_{n}^{\dagger}(\hatd_{n})\right\} $
where the states $p,p\'$ can be in any region (lead or central) of
the total system. In doing so, the stochastic Hamiltonian can be more
compactly expressed, 
\begin{align}
\hatH^{\kappa}(z) & =\sum_{pp\'}h_{pp\'}(z)\hata_{p}^{\dagger}\hata_{p\'}\label{eq:hkappa}\\
 & =\begin{cases}
\ \hatH^{\pm}(t)=\hatH^{0}(t)+\hatH_{el-ph}^{\pm}(t),\quad & \textnormal{on}\ \kappa^{\pm}\\
\ \mathcal{\hatH}(\tau)=\mathcal{\hatH}^{0}+\overline{{H}}_{el-ph}(\tau), & \textnormal{on}\ \kappa^{M}.
\end{cases}\label{eq:Hkappa}
\end{align}
Writing the Hamiltonian in this way makes it possible to construct
blocks of the matrix $h_{pp\'}(t)$ projected onto the lead and central
region subspaces for any combination of times on the contour, 
\begin{align}
\left\{ h_{pp\'}(z)\right\} =\begin{cases}
h_{LL}(t)=\{h_{ij}(t)\}=\left\{ \delta_{ij}\left[\epsilon_{i}+V_{L}(t)\right]\right\} ,\,\,\text{where}\,\,i,j\in L\,\,\text{and}\,\,t\in\kappa_{\pm}\\
h_{LL}^{M}(\tau)=\{h_{ij}^{M}(\tau)\}=\{\delta_{ij}\epsilon_{i}^{M}\},\ \textnormal{where}\ i,j\in L,\,\,\,\tau\in\kappa^{M}\\
h_{LC}=\{h_{in}\}=\{T_{in}\},\,\,\text{where}\,\,i\in L,\,\,n\in C\,\,\,\text{and}\,\,z\in\text{\ensuremath{\kappa}=\ensuremath{\kappa_{\pm}\cup\kappa^{M}}}\\
h_{CC}^{\pm}(t)=\{h_{nm}^{\pm}(t)\}=\left\{ T_{nm}+w_{nm}^{\pm}(t)\right\} ,\,\,\text{where}\,\,n,m\in C\,\,\text{and}\,\,t\in\kappa_{\pm}\\
\overline{h}_{CC}(\tau)=\{\overline{h}_{nm}(\tau)\}=\left\{ T_{nm}+\overline{w}_{nm}(\tau)\right\} ,\ \textnormal{where}\ n,m\in C,\,\,\,\tau\in\kappa^{M}.
\end{cases}\label{eq:hpp'kappa}
\end{align}
Above, $\epsilon_{i}^{M}=\epsilon_{i}-\mu_{L}$ and the stochastic
part of $h_{CC}(z)$ is contained within the unravelling matrix $w_{CC}(z)$
which is given by:
\begin{gather}
w_{CC}^{\pm}(t)=-\sum_{A\alpha}\left[\eta_{A\alpha}(t)\pm\frac{\hbar}{2}\nu_{A\alpha}(t)\right]\mathcal{V}_{CC}^{A\alpha}\quad\textnormal{on}\ \kappa^{\pm}\label{eq:W}\\
\overline{w}_{CC}(\tau)=-\sum_{A\alpha}\overline{\mu}_{A\alpha}(\tau)\mathcal{V}_{CC}^{A\alpha}\quad\textnormal{on}\ \kappa^{M}.\label{eq:WM}
\end{gather}
We take a moment to emphasise this notation. The labels $\pm$ indicate
the presence of stochasticity on the upper and lower horizontal branches,
while the over-bar indicates stochasticity on the vertical branch.
This is distinct from the label $M$ for the regular Matsubara branch
which does not have any stochasticity associated with it.

The physical density matrix is then recovered by the stochastic average,
\begin{align}
\rhohat(\overline{t})=\mathbb{N}\langle\widetilde{\rho}(\overline{t})\rangle_{\xi\overline{\xi}},
\end{align}
where $\xi$ and $\overline{\xi}$ represent the noises in real and
imaginary time, respectively (Eq. \eqref{eq:xi} in Appendix \ref{A:influence})
and
\begin{align}
\mathbb{N}=\langle\bar{\rho}(\beta\hbar)\rangle_{\xi\overline{\xi}}^{-1}=\langle\overline{\mathcal{U}}(\beta\hbar,0)\rangle_{\xi\overline{\xi}}^{-1}\label{eq:Nrhobar}
\end{align}
is a normalisation factor which ensures that $\Tr\left[\rhohat\right]$=1,
and is needed because of the non-Hermicity of the stochastic dynamics.
The noises themselves have non-trivial correlation functions (Eqs.
\eqref{eq:etaeta}-\eqref{eq:mumu} in Appendix \ref{A:influence}),
and are in no way arbitrary or introduced \textit{ad hoc}. Rather,
they are related to the actual phonon dynamics and require knowledge
of the phonon eigenstates. Note that this is the only place where
actual information of the phonons appears.

\section{The Three-Time Green's Function}

\subsection{Additional Branch Dependence}

\label{sec:branchdependence} To justify an introduction of the three-time
Green's function, let us calculate the electronic population matrix
at time $\overline{t}$, 
\begin{align}
P_{pp\'}(\overline{t}) & =\Tr\left[\rhohat(\overline{t})\hata_{p}^{\dagger}\hata_{p\'}\right].
\end{align}
Using the stochastic unravelling procedure laid out in Section \ref{sec:HS},
the population matrix is unravelled as 
\begin{align}
P_{pp\'}(\overline{t}) & =\mathbb{N}\left\langle \Tr\left[\overline{\mathcal{U}}(\beta\hbar,0)\hatU^{-}(t_{0}^{-},\overline{t})\hata_{p}^{\dagger}\hata_{p\'}\hatU^{+}(\overline{t},t_{0}^{+})\right]\right\rangle _{\xi\overline{\xi}}\label{eq:Punravelling}\\
 & \equiv\langle\widetilde{P}_{pp\'}(\overline{t})\rangle_{\xi\overline{\xi}},
\end{align}
where the tilde on $\widetilde{P}$ again indicates that it is a stochastic
quantity; it corresponds to the population calculated for the Hamiltonian
(\ref{eq:hpp'kappa}) in which the atomic displacements in the electron-phonon
coupling were replaced with the noises and hence corresponds to a
particular stochastic realisation.

In order to calculate the populations $\widetilde{P}_{pp'}(\overline{t})$
appearing in Eq. \eqref{eq:Punravelling}, we define the three-time
NEGF, 
\begin{align}
G_{pp\'}(z,z\'\vert\overline{t}) & =-\frac{i}{\hbar}\frac{1}{\Pi(\overline{t})}\begin{cases}
\Tr\left[\hatU(t_{0}^{-}-i\beta\hbar,z)\hata_{p}\hatU(z,z\')\hata_{p\'}^{\dagger}\hatU(z\',t_{0}^{+})\right],\qquad & \textnormal{for}\ z>z\'\ \textnormal{w.r.t}\ \hatT_{\kappa}\\
-\Tr\left[\hatU(t_{0}^{-}-i\beta\hbar,z\')\hata_{p\'}^{\dagger}\hatU(z\',z)\hata_{p}\hatU(z,t_{0}^{+})\right], & \textnormal{for}\ z<z\'\ \textnormal{w.r.t}\ \hatT_{\kappa}
\end{cases},\label{eq:threetimeG}
\end{align}
where the function $\Pi(\overline{t})$ (given below in Eq. \eqref{eq:Pi1})
is introduced to ensure that the Green's function still satisfies
the regular equations of motion, 
\begin{gather}
i\hbar\partial_{z}G_{pp\'}(z,z\'\vert\overline{t})=\delta(z-z\')\delta_{pq}+\sum_{q}h_{pq}(z)G_{qp\'}(z,z\'\vert\overline{t})\label{eq:eqmG}\\
-i\hbar\partial_{z'}G_{pp\'}(z,z\'\vert\overline{t})=\delta(z-z\')\delta_{pq}+\sum_{q}G_{pq}(z,z\'\vert\overline{t})h_{qp\'}(z),
\end{gather}
and the $h_{pq}(z)$ are from Eq. \eqref{eq:hkappa}. Here, the third
time $\overline{t}$ has been introduced, and is referred to as the
\textit{observation time} (see Figure \ref{fig:contour}), and the
full time evolution operator across the contour $\kappa$ is required,
defined by the equations of motion ($z\in\kappa$), 
\begin{gather}
i\hbar\partial_{t}\hatU(t,z)=\hatH^{\pm}(t)\hatU(t,z)\\
i\hbar\partial_{t}\hatU(z,t)=-\hatU(z,t)\hatH^{\pm}(t)\label{eq:dtUzz}\\
-\hbar\partial_{\tau}\hatU(t_{0}^{-}-i\tau,z)=\mathcal{\hat{H}}(\tau)\hatU(t_{0}^{-}-i\tau,z)\\
-\hbar\partial_{\tau}\hatU(z,t_{0}^{-}-i\tau)=-\hatU(z,t_{0}^{-}-i\tau)\mathcal{\hat{H}}(\tau),\label{eq:dtauUzz}
\end{gather}
which have the solution, 
\begin{equation}
\hatU(z,z\')=\hatT_{\kappa}\exp\left\{ -\frac{i}{\hbar}\int_{z\'}^{z}dz_{1}\hatH^{\kappa}(z_{1})\right\} .\label{eq:U1}
\end{equation}
For instance, when both times are on the horizontal branches $\kappa^{\pm}$,
we arrive at $\hatU^{\pm}(t,t\')$, where $\hatT_{\kappa}\rightarrow\hatT_{\pm}$
and $\hatH^{\kappa}\rightarrow\hatH^{\pm}(t)$, while when both time
arguments are on the vertical track $\kappa^{M}$, we have $\overline{\mathcal{U}}(\tau,\tau\')$,
in which $\hatT_{\kappa}\rightarrow\hatT_{M}$ and $\hatH^{\kappa}\rightarrow\mathcal{\hatH}(\tau)$.
When both times $z$ and $z'$ belong to different tracks on the contour
$\kappa$, an integration over the appropriate part of the contour
from $z'$ to $z$ is implied, with appropriate Hamiltonian on each
part of the relevant tracks.

The dependence on the observation time $\overline{t}$ which appears
as the third time is a subtle point that must be emphasized. Since
the Hamiltonians on the upper and lower branches are now unique, there
is no cancellation for regions of the contour which would normally
be shared between them, for example the deterministic dynamics of
the regular Green's function from $t\'$ to $\overline{t}$ on $\kappa^{+}$
would annul the dynamics from $\overline{t}$ to $t\'$ on $\kappa^{-}$.
Instead, the full dynamics of the three-time Green's function on $\kappa^{+}$
and $\kappa^{-}$ must be considered independently, with $\overline{t}$
parameterizing the dynamics up to some upper time limit $t_{max}$.
This means that the Heisenberg representation cannot be used in the
normal way \cite{PhysRevB.101.165408}, since $\hatH^{+}\neq\hatH^{-}$
and $\hatU^{+}$ and $\hatU^{-}$ no longer form a Hermitian conjugate
pair; hence the need for the introduction of this new kind of three-time
NEGF.

The stochastic population matrix is then obtained via relation to
a particular component of this NEGF, 
\begin{equation}
\widetilde{P}(\overline{t})=-i\hbar\mathbb{N}\Pi(\overline{t})G^{<}(\overline{t}^{+},\overline{t}^{-}\vert\overline{t})=-i\hbar\mathbb{N}\Pi(\overline{t})G^{+-}(\overline{t},\overline{t}\vert\overline{t}),\label{eq:popMatrix}
\end{equation}
where $\overline{t}^{-}$ and $\overline{t}^{+}$ are the observation
times taken on the lower and upper horizontal branches, respectively,
with $\overline{t}^{+}$ being just before and $\overline{t}^{-}$
just after the actual observation time $\overline{t}$ on the contour.
The calculation requires the specific lesser Green's function $G^{+-}$
(see \cite{PhysRevB.101.165408}) with the first argument on the upper
and the second on the lower horizontal branches, but when both tend
to the observation time. Hence, the task becomes calculating the desired
blocks of this lesser component of the Green's function Eq. \eqref{eq:threetimeG}
such as $G_{LL}$, $G_{CC}$, etc. when all three times are equal
to the observation time.

At the same time, one needs to calculate the prefactor $\mathbb{N}\Pi(\overline{t})$.
In the rest of this section we shall consider how to calculate this
prefactor, while the method of calculating the three-time Green's
function will be considered in the next section.

The function $\Pi(\overline{t})$ takes the form 
\begin{equation}
\Pi(\overline{t})=\Tr\left[\hatU(t_{0}^{-}-i\beta\hbar,t_{0}^{+})\right]
\end{equation}

\begin{equation}
=\Tr\left[\hatU(t_{0}^{-}-i\beta\hbar,t_{0}^{-})\hatU(t_{0}^{-},\overline{t})\hatU(\overline{t},t_{0}^{+})\right]\label{eq:Pi1}
\end{equation}
where the $\overline{t}$ dependence has been made explicit by showing
the shared time in the propagators. Hence, the trace in Eq. \eqref{eq:Pi1}
is calculated over the complete propagation along the contour, from
the initial time $t_{0}^{+}$ to the final $t_{0}^{-}-i\beta\hbar$
passing through the \textit{observation time} $\overline{t}$ on the
way. Differentiating $\Pi(\overline{t})$ with respect to $\overline{t}$
yields 
\begin{equation}
i\hbar\partial_{\overline{t}}\Pi(\overline{t})=\hbar\left[i\hbar\Pi(\overline{t})\right]\sum_{A\alpha}\nu_{A\alpha}(\overline{t})\,\text{tr}\left[\mathcal{V}_{}^{A\alpha}G_{CC}^{+-}(\overline{t},\overline{t}\vert\overline{t})\right],
\end{equation}
which has the formal solution: 
\begin{align}
\Pi(\overline{t})=\Pi_{0}\exp\left\{ \hbar\int_{t_{0}}^{\overline{t}}dt\ \sum_{A\alpha}\nu_{A\alpha}(t)\:\text{tr}\left[\mathcal{V}^{A\alpha}G_{CC}^{+-}(t,t\vert t)\right]\right\} ,\label{eq:Pi2}
\end{align}
where we distinguish between the trace of a matrix and the quantum-mechanical
trace by writing the former using small letters. Since the electron-phonon
coupling matrix $\mathcal{V}$ is only non-zero for electronic states
in the central region, only the central region Green's function appears
in the trace.

Here, $\Pi_{0}=\Pi(t_{0}^{-})$ is the initial value of $\Pi$ when
$\overline{t}=t_{0}^{-}=t_{0}^{+}$, that is, when there are no horizontal
branches and only the Matsubara branch remains. To calculate $\Pi_{0}$,
it is convenient to introduce yet another three-time Green's function
which exists only on the vertical branch which we shall refer to as
the \textit{thermal} Green's function:
\begin{align}
\overline{G}_{pp\'}(\tau,\tau\'\vert\overline{\tau}) & =-\frac{1}{\hbar}\frac{1}{\overline{\Pi}(\overline{\tau})}\begin{cases}
\Tr\left[\barU(t_{0}^{-},\tau)\hata_{p}\barU(\tau,\tau\')\hata_{p\'}^{\dagger}\barU(\tau\',t_{0}^{-}-i\overline{\tau})\right]\qquad & \textnormal{for}\ \tau>\tau\',\ \tau,\tau\'\in(0,\overline{\tau})\\
-\Tr\left[\barU(t_{0}^{-},\tau\')\hata_{p\'}^{\dagger}\barU(\tau\',\tau)\hata_{p}\barU(\tau,t_{0}^{-}-i\overline{\tau})\right] & \textnormal{for}\ \tau<\tau,\ \tau,\tau\'\in(0,\overline{\tau})
\end{cases}\label{eq:3-time-thermal-GF}
\end{align}
which involves the imaginary time propagator Eq. \eqref{eq:Ubar}
defined on the vertical track only, $\overline{\mathcal{U}}(\tau,\tau')$.
Note that this $\overline{G}$ is not the same as the normal Matsubara
component of the regular Green's function whose arguments $\tau,\tau\'$
are defined on the entire vertical branch $\kappa^{M}$, $\tau,\tau\'\in(0,\beta\hbar)$.
Instead, the thermal three-time Green's function contains $0\le\overline{\tau}\le\beta\hbar$
as the third imaginary time, and hence is defined on the subbranch
within $\kappa^{M}$ from $0$ to $\overline{\tau}$ only. Since $\overline{\tau}$
is responsible for extending the subbranch up to $\kappa^{M}$, it
is responsible for the thermalization of the total system including
phonons into the canonical equilibrium state, and shall be referred
to as the \textit{preparation} time.

With these definitions, $\Pi_{0}$ is related to the imaginary time
propagator for $\bar{\tau}=\beta\hbar$, 
\begin{align}
\Pi_{0} & =\Tr\left[\hatU(t_{0}^{-}-i\beta\hbar,t_{0}^{-})\hatU(t_{0}^{-},t_{0}^{+})\right]\\
 & =\Tr\left[\overline{\mathcal{U}}(\beta\hbar,0)\right],
\end{align}
since $\hatU(t_{0}^{-},t_{0}^{+})=1$. This leads to the definition
of a similar function to Eq. \eqref{eq:Pi1}, only this time on the
vertical branch, 
\begin{align}
\overline{\Pi}(\overline{\tau}) & =\Tr\left[\overline{\mathcal{U}}(\overline{\tau},0)\right],\label{eq:Pibardef}
\end{align}
so that 
\begin{equation}
\Pi_{0}=\overline{\Pi}(\beta\hbar).\label{eq:Pi-0-via-beta-h}
\end{equation}
The auxiliary function $\overline{\Pi}(\overline{\tau})$ has the
equation of motion, 
\begin{align}
-\hbar\partial_{\overline{\tau}}\overline{\Pi}(\overline{\tau}) & =\hbar\overline{\Pi}(\overline{\tau})\,\text{tr}\left[\overline{h}(\overline{\tau})\overline{G}^{<}(0,0^{+}\vert\overline{\tau})\right],\label{eq:Pibareqm}
\end{align}
in which the lesser component of the three-time thermal Green's function
appears. Note that since $\overline{G}^{<}$ is defined only on the
vertical branch, both its arguments belong to the same branch making
it clearly determined. Integrating Eq. \eqref{eq:Pibareqm} yields
\begin{equation}
\overline{\Pi}(\overline{\tau})=\overline{\Pi}_{0}\exp\left\{ -\int_{0}^{\overline{\tau}}d\tau\,\text{tr}\left[\overline{h}(\tau)\overline{G}^{<}(0,0^{+}\vert\tau)\right]\right\} ,\label{eq:Pibarintermediary}
\end{equation}
where $\overline{\Pi}_{0}=\overline{\Pi}(0)=\Tr\left[\overline{U}(0,0)\right]=1$.

Crucially, $\overline{\Pi}(\overline{\tau})$ depends on the entire
three-time thermal Green's function across all regions of the junction.
Expanding the trace using block notation for the matrices of each
region of the junction, each block of $\overline{G}$ can be expressed
in terms of the central region $\overline{G}_{CC}$ by first introducing
the three-time isolated lead Green's function $\overline{g}_{LL}^{0}(\tau,\tau\'\vert\overline{\tau})$,
defined on the subbranch of $\kappa^{M}$ between $0$ and $\overline{\tau}$.
Its equation of motion (in obvious symbolic notation) is 
\begin{align}
-(\hbar\partial+h_{LL}^{M})\overline{g}_{LL}^{0}=\delta,\label{eq:overlineg_LL^0}
\end{align}
which has the solution 
\begin{align}
\overline{g}_{ij}(\tau,\tau\'\vert\overline{\tau})=-\frac{1}{\hbar}\delta_{ij}e^{-\epsilon_{i}^{M}(\tau-\tau\')/\hbar}\left[\Theta(\tau-\tau\')\left[1-\overline{f}(\epsilon_{i}^{M}\vert\overline{\tau})\right]-\Theta(\tau\'-\tau)\overline{f}(\epsilon_{i}^{M}\vert\overline{\tau})\right],\label{eq:gbar}
\end{align}
where $\overline{f}(\epsilon\vert\overline{\tau})$ is just the ordinary
Fermi function but with $\beta\hbar$ replaced by $\overline{\tau}$.
Note that $\overline{g}_{LL}^{0}$ is not a stochastic Green's function;
the $\overline{\tau}$ dependence comes solely from these modified
Fermi functions.

The equation of motion for $\overline{G}_{LC}$ is 
\begin{equation}
-\left(\hbar\partial+{h}_{LL}^{M}\right)\Gbar_{LC}={h}_{LC}\Gbar_{CC},
\end{equation}
which by inverting Eq. \eqref{eq:overlineg_LL^0} becomes 
\begin{equation}
\Gbar_{LC}=\overline{g}_{LL}^{0}{h}_{LC}\Gbar_{CC},
\end{equation}
and similarly for $\Gbar_{CL}$, 
\begin{equation}
\Gbar_{CL}=\Gbar_{CC}{h}_{CL}\overline{g}_{LL}^{0}.\label{eq:Gbar_CL}
\end{equation}
Then finally for $\Gbar_{LL}$, 
\begin{align}
-\left(\hbar\partial+{h}_{LL}^{M}\right)\Gbar_{LL}=\delta_{L}+{h}_{LC}\Gbar_{CL},
\end{align}
applying Eqs. \eqref{eq:overlineg_LL^0} and \eqref{eq:Gbar_CL} gives
\begin{equation}
\Gbar_{LL}=\overline{g}_{LL}^{0}+\overline{g}_{LL}^{0}{h}_{LC}\Gbar_{CC}{h}_{CL}\overline{g}_{LL}^{0}.
\end{equation}
Substituting these into Eq. \eqref{eq:Pibarintermediary} and using
cyclic permutations of the trace and the fact that different leads
do not interact, $h_{LL'}=0$ for $L\ne L'$, we obtain 
\begin{align}
\overline{\Pi}(\overline{\tau}) & =\exp\left\{ -\int_{0}^{\overline{\tau}}d\tau\,\bigg(\text{tr}\left[\overline{h}_{CC}(\tau)\overline{G}_{CC}^{<}(0,0^{+}\vert\tau)\right]+\sum_{L}\text{tr}\left[h_{LL}^{M}\overline{g}_{LL}^{0<}(0,0^{+}\vert\tau)\right]\right.\nonumber \\
 & \qquad\left.\left.+\int_{0}^{\tau}d\tau_{1}\,\text{tr}\left[\overline{\Sigma}_{CC}^{<}(0,\tau_{1}\vert\tau)\overline{G}_{CC}^{>}(\tau_{1},0^{+}\vert\tau)+\overline{G}_{CC}^{<}(0,\tau_{1}\vert\tau)\overline{\Sigma}_{CC}^{>}(\tau_{1},0^{+}\vert\tau)\right]\right.\right.\nonumber \\
 & \qquad\left.\left.+\int_{0}^{\tau}d\tau_{1}\int_{0}^{\tau}d\tau_{2}\,\text{tr}\left[\overline{\Lambda}_{CC}(\tau_{2},\tau_{1}\vert\tau)\overline{G}_{CC}(\tau_{1},\tau_{2}\vert\tau)\right]\right)\right\} ,\label{eq:Pibarintermediary2}
\end{align}
where $\overline{h}_{CC}(\tau)$ contains the noises $\{\overline{\mu}_{A\alpha}(\tau)\}$
and two new imaginary time self-energies $\overline{\Sigma}_{CC}$
and $\overline{\Lambda}_{CC}$ have been defined:
\begin{gather}
\overline{\Sigma}_{CC}(\tau,\tau\'\vert\overline{\tau})=\sum_{L}\overline{\Sigma}_{CLC}(\tau,\tau\'\vert\overline{\tau})=\sum_{L}h_{CL}\overline{g}_{LL}^{0}(\tau,\tau\'\vert\overline{\tau})h_{LC}\\
\overline{\Lambda}_{CC}(\tau,\tau\'\vert\overline{\tau})=\sum_{L}\overline{\Lambda}_{CLC}(\tau,\tau\'\vert\overline{\tau})=\sum_{L}h_{CL}\overline{g}_{LL}^{0>}(\tau,0^{+}\vert\overline{\tau})h_{LL}^{M}\overline{g}_{LL}^{0<}(0,\tau\'\vert\overline{\tau})h_{LC}.
\end{gather}

Returning to Eq. \eqref{eq:Pibarintermediary2}, the presence of the
term containing $\sum_{L}\text{tr}\left[h_{LL}^{M}\overline{g}_{LL}^{0<}(0,0^{+}\vert\tau)\right]$
is concerning at first glance as the semi-infinite nature of the leads
means it will in general be infinity. Since it does not depend on
the noises, it can be taken outside of any stochastic averages as
a noise-independent pre-factor, 
\begin{align}
Y(\overline{\tau})=\exp\left\{ -\int_{0}^{\overline{\tau}}d\tau\,\sum_{L}\text{tr}\left[h_{LL}^{M}\overline{g}_{LL}^{0<}(0,0^{+}\vert\tau)\right]\right\} ,
\end{align}
so that $\overline{\Pi}(\overline{\tau})=Y(\overline{\tau})\overline{\Pi}_{1}(\overline{\tau}),$
where $\overline{\Pi}_{1}$ is the remaining noise-dependent part.
When calculating the population matrix (see Eqs. \eqref{eq:popMatrix},
\eqref{eq:Pi2} and \eqref{eq:Pi-0-via-beta-h}), this infinite prefactor
appears in two places. First, it appears in the normalisation constant
$\mathbb{N}$ in Eq. \eqref{eq:popMatrix}, defined in Eq. \eqref{eq:Nrhobar},
which takes the value 
\[
\mathbb{N}=\langle\overline{\Pi}(\beta\hbar)\rangle_{\xi\overline{\xi}}^{-1}=Y(\beta\hbar)^{-1}\langle\overline{\Pi}_{1}(\beta\hbar)\rangle_{\xi\overline{\xi}}^{-1},
\]
since $\overline{\rho}(\tau)=\overline{\mathcal{U}}(\tau,0)$. And
second, it appears in the factor $\Pi_{0}=\overline{\Pi}(\beta\hbar)$
within the definition of $\Pi(\overline{t})$. Hence, substituting
these results into Eq. \eqref{eq:popMatrix}, the physical population
matrix after stochastic averaging becomes 
\begin{align}
P_{pp\'}(\overline{t})=-i\hbar\frac{\langle\Pi_{1}(\overline{t})G^{+-}(\overline{t},\overline{t}\vert\overline{t})\rangle_{\xi\overline{\xi}}}{\langle\overline{\Pi}_{1}(\beta\hbar)\rangle_{\xi\overline{\xi}}},
\end{align}
where $\Pi_{1}(\overline{t})$ is simply $\Pi(\overline{t})$ but
with the factor of $Y(\beta\hbar)$ removed from $\Pi_{0}$. The factors
of $Y(\beta\hbar)$ thus appear in both the numerator and denominator
and cancels out, so it is convenient to redefine $\overline{\Pi}(\overline{\tau})$
to only include the stochastic part $\overline{\Pi}_{1}(\overline{\tau})$,
as well as redefining $\Pi(\overline{t})$ to only include $\overline{\Pi}_{1}(\beta\hbar)$
in $\Pi_{0}$. For completeness, the final form of $\overline{\Pi}(\overline{\tau})$
is 
\begin{align}
\overline{\Pi}(\overline{\tau}) & =\exp\left\{ -\int_{0}^{\overline{\tau}}d\tau\,\bigg(\text{tr}\left[\overline{h}_{CC}(\tau)\overline{G}_{CC}^{<}(0,0^{+}\vert\tau)\right]\right.\nonumber \\
 & \qquad\left.\left.+\int_{0}^{\tau}d\tau_{1}\,\text{tr}\left[\overline{\Sigma}_{CC}^{<}(0,\tau_{1}\vert\tau)\overline{G}_{CC}^{>}(\tau_{1},0^{+}\vert\tau)+\overline{G}_{CC}^{<}(0,\tau_{1}\vert\tau)\overline{\Sigma}_{CC}^{>}(\tau_{1},0^{+}\vert\tau)\right]\right.\right.\nonumber \\
 & \qquad\left.\left.+\int_{0}^{\tau}d\tau_{1}\int_{0}^{\tau}d\tau_{2}\,\text{tr}\left[\overline{\Lambda}_{CC}(\tau_{2},\tau_{1}\vert\tau)\overline{G}_{CC}(\tau_{1},\tau_{2}\vert\tau)\right]\right)\right\} ,\label{eq:Pibar}
\end{align}
which can be calculated as long as the three-time thermal Green's
function in the central region $\overline{G}_{CC}$ is known.

\subsection{Series Expansion in $W$}

\label{sec:W} The equations of motion for blocks of the three-time
NEGF in different regions of the junction (in obvious symbolic notation)
are: 
\begin{gather}
(i\hbar\partial-h_{CC})G_{CC}=\delta_{C}+\sum_{L}h_{CL}G_{LC}\label{eq:G_CC}\\
(i\hbar\partial-h_{CC})G_{CL}=h_{LL}G_{LL}\label{eq:G_CL}\\
(i\hbar\partial-h_{LL})G_{LC}=h_{LC}G_{CC}\label{eq:G_LC}\\
(i\hbar\partial-h_{LL})G_{LL}=\delta_{L}+h_{LC}G_{CL}.\label{eq:G_LL}
\end{gather}
Here, $h_{CC}(z)$ can be split into the phonon-free part $h_{CC}^{0}=\{T_{nm}\}$
and the unravelled part from Eqs. \eqref{eq:W} and \eqref{eq:WM}
which contain the noises. Substituting Eq. \eqref{eq:G_LC} into Eq.
\eqref{eq:G_CC} yields 
\begin{align}
(i\hbar\partial-h_{CC}^{0})G_{CC}=\delta_{C}+W_{CC}G_{CC}+\Sigma_{CC}G_{CC},\label{eq:G_CCeqm}
\end{align}
where the general two-time unravelling matrix $W_{CC}(z,z\')$ has
been introduced,
\begin{align}
W_{CC}(z,z\')=\begin{cases}
\ \delta(t-t\')w_{CC}^{\pm}(t) & z,z\'=t,t\'\in\kappa^{\pm}\\
\ \delta(\tau-\tau\')\overline{w}_{CC}(\tau) & z,z\'=\tau,\tau\'\in\kappa^{M}\\
\ 0 & \textnormal{otherwise,}
\end{cases}\label{eq:W_CC-2-times}
\end{align}
in place of Eqs. \eqref{eq:W} and \eqref{eq:WM} for ease of notation
when appearing in contour integrals, and 
\begin{align}
\Sigma_{CC}(z,z\')=\sum_{L}h_{CL}g_{LL}^{0}(z,z\')h_{LC},
\end{align}
is the regular embedding self-energy containing the Green's function
of the isolated leads $g_{LL}^{0}$ under the bias. Similarly, the
equation of motion for the phonon-free Green's function in which phonons
are not accounted for (with the same bias) is 
\begin{align}
(i\hbar\partial-h_{CC}^{0})G_{CC}^{0}=\delta_{C}+\Sigma_{CC}G_{CC}^{0},
\end{align}
Comparing this with Eq. \eqref{eq:G_CCeqm} one obtains a self-consistent
Dyson-like formal equation for $G_{CC}$, 
\begin{equation}
G_{CC}=G_{CC}^{0}+G_{CC}^{0}W_{CC}G_{CC},\label{eq:G=00003DG0+G0WG}
\end{equation}
which can be used to generate a Born-like series expansion of the
full three-time NEGF in terms of the noises and the phonon-free Green's
function, which in the symbolic form reads:
\begin{align}
G_{CC}=G_{CC}^{0}+G_{CC}^{0}W_{CC}G_{CC}^{0}+G_{CC}^{0}W_{CC}G_{CC}^{0}W_{CC}G_{CC}^{0}+\ldots.
\end{align}

Writing the times in Eq. \eqref{eq:G=00003DG0+G0WG} explicitly, 
\begin{align}
G_{CC}(z,z\'\vert\overline{t})=G_{CC}^{0}(z,z\')+\int_{\kappa}dz_{1}dz_{2}G_{CC}^{0}(z,z_{1})W_{CC}(z_{1},z_{2})G_{CC}(z_{2},z\'\vert\overline{t}),\label{eq:explicitDyson}
\end{align}
the integrals over $\kappa$ must be taken with respect to the generalized
Langreth rules\cite{PhysRevB.101.165408} for the specific component
of $G_{CC}$ of interest. As an example, an expansion for the lesser
component $G^{+-}(z,z'|\overline{t})$ is presented in Appendix \ref{A:langreth}.

Eq. \eqref{eq:explicitDyson} generates a perturbative expansion with
respect to the unravelling matrix $W_{CC}$. The expansion for the
purposes of the calculation must be truncated at certain order with
respect to the unravelling matrix $W_{CC}$ that is linear with respect
to the noises.Note that this also requires knowledge of the components
of the phonon-free Green's function $G_{CC}^{0}$ for a variable bias;
these expressions are readily available, e.g., in the wide band approximation
\cite{PhysRevB.91.125433}. Hence, the three-time Green's function
can be written explicitly in a Born-like series with respect to the
electrons-only Green's function of the junction (which is assumed
known) and the noises, so that $G_{CC}$ is expressed as a power series
with respect to the noises.

In addition to the three-time Green's function $G_{CC}$, we also
need a working expression for the thermal three-time Green's function;
this can also be expanded in a similar fashion, 
\begin{align}
\Gbar_{CC}(\tau,\tau\'\vert\overline{\tau})=G_{CC}^{0M}(\tau,\tau\')+\int_{0}^{\overline{\tau}}d\tau_{1}G_{CC}^{0M}(\tau,\tau_{1})\overline{w}_{CC}(\tau_{1})\Gbar_{CC}(\tau_{1},\tau\'\vert\overline{\tau}),
\end{align}
where this time generalised Langreth rules are not required as the
integration is performed over the subbranch of $\kappa^{M}$ from
0 up to the preparation time $\overline{\tau}$.

\subsection{The Current}

\label{sec:current} Now that the central region three-time Green's
function can be calculated, any block of the Green's function can
be found. For example, the Green's function for the $L^{th}$ lead,
\begin{equation}
G_{LL}={g}_{LL}^{0}+{g}_{LL}^{0}{h}_{LC}G_{CC}{h}_{CL}{g}_{LL}^{0}.\label{eq:GLL}
\end{equation}
appears in the unravelled number operator which is just the trace
of the population matrix Eq. \eqref{eq:popMatrix} of the lead,
\begin{equation}
\widetilde{N}_{L}(\overline{t})=-i\hbar\mathbb{N}\Pi(\overline{t})\,\text{tr}\left[G_{LL}^{+-}(\overline{t},\overline{t}\vert\overline{t})\right]
\end{equation}
\begin{equation}
=-i\hbar\mathbb{N}\Pi(\overline{t})\left(\text{tr}\left[g_{LL}^{0<}(\overline{t},\overline{t})\right]+\int_{\kappa}dz_{1}dz_{2}\,\text{tr}\left[g_{LL}^{0}\left(\overline{t}^{+},z_{1}\right)h_{LC}G_{CC}(z_{1},z_{2}\vert\overline{t})h_{CL}g_{LL}^{0}\left(z_{2},\overline{t}^{-}\right)\right]\right)\label{eq:NumberL}
\end{equation}
where the double contour integral is again taken with respect to the
generalized Langreth rules, and the expression in the square brackets
is understood as the $+-$ projection; this is stated explicitly by
the time arguments of the $g_{LL}^{0}$, the first one containing
$\overline{t}^{+}$ and the second $\overline{t}^{-}$. Now, using
cyclic invariance of the trace, we can rearrange:

\begin{equation}
\widetilde{N}_{L}(\overline{t})=-i\hbar\mathbb{N}\Pi(\overline{t})\left(\text{tr}\left[g_{LL}^{0<}(\overline{t},\overline{t})\right]+\int_{\kappa}dz_{1}dz_{2}\,\text{tr}\left[\Upsilon_{CLC}(z_{2},z_{1})G_{CC}(z_{1},z_{2}\vert\overline{t})\right]\right)\label{eq:lead-L-population}
\end{equation}
having introduced a new self-energy with components
\begin{align}
\Upsilon_{CLC}^{\gamma\gamma\'}(z,z\')=h_{CL}g_{LL}^{0\gamma}(z,\overline{t}^{-})g_{LL}^{0\gamma\'}(\overline{t}^{+},z\')h_{LC},\label{eq:Y-self-energy-for-current}
\end{align}
where the superscripts $\gamma,\gamma\'$ specify the projections
of the isolated lead Green's functions onto the different branches
of the contour, as well as their ordering with respect to $\hatT_{\kappa}$.
Note that this self-energy depends on only two times on the contour
$z$ and $z'$, rather than all four of the arguments which appear
on the right hand side as the inner two times are explicitly set to
the observation time $\overline{t}^{\pm}$ of the indicated branch
(upper and lower).

The first term in Eq. \eqref{eq:lead-L-population} involving the
isolated lead $g_{LL}^{0<}$ will be infinite due to the trace over
the orbitals in a semi-infinite lead, where for equal arguments, $g_{LL}^{0<}$
is time-independent and is just the Fermi function (Eq. \eqref{eq:g<}).
However, it can be shown that this term, after stochastic averaging,
represents the total number of electrons in the isolated lead which
is a time-independent quantity (see Appendix \ref{A:Girsanov}). Hence,
only the second term in Eq. \eqref{eq:lead-L-population} is time-dependent
and responsible for the current.

Hence, differentiating the second term in the number operator of the
lead and multiplying by the electron charge, $-e$, we obtain the
current (note that the derivative is a linear operator which commutes
with the stochastic average): 
\begin{align}
\hat{J}_{L}(\overline{t}) & =\frac{i\hbar e}{\big\langle\overline{\Pi}(\beta\hbar)\big\rangle_{\xi\overline{\xi}}}\partial_{\overline{t}}\bigg\langle\Pi(\overline{t})\int_{\kappa}dz_{1}dz_{2}\,\text{tr}\left[\Upsilon_{CLC}(z_{2},z_{1})G_{CC}(z_{1},z_{2}\vert\overline{t})\right]\bigg\rangle_{\xi\overline{\xi}}\,.\label{eq:current}
\end{align}
Note that the double contour integral over both the inner and outer
times of $\Upsilon_{CLC}(z_{2},z_{1})G_{CC}(z_{1},z_{2}\vert\overline{t})$
cannot be written by applying the generalized Langreth rules alone.
Instead, all possible combinations of times on the contour with all
possible time orderings must be seperately considered, introducing
the need for the $\gamma,\gamma\'$ indices in Eq. \eqref{eq:Y-self-energy-for-current}.
An explicit expression for the trace in the integrand via various
components of the Green's function is given in Appendix \ref{A:langreth},
while the derivation of the components $\Upsilon_{CLC}^{\gamma\gamma\'}$
are presented in Appendix \ref{A:selfenergies}.

\section{Discussion}

\label{sec:discussion} Having derived the three-time NEGF and it's
associated functions, we should review how it is different from previous
approaches. We began by including a phonon bath in the central region
Hamiltonian using phonon displacements from equilibrium where the
coupling between the bath and central region electrons is linear with
respect to these displacements but arbitrary with respect to electrons.
Using path integrals to integrate out the effect of the phonons on
the electronic system, we were able to apply a HS transformation to
the influence functional which exactly removed the phonon degrees
of freedom all together, replacing them with three sets of coloured
Gaussian noises $\{\eta_{A\alpha}(t)\}$, $\{\nu_{A\alpha}(t)\}$
and $\{\overline{\mu}_{A\alpha}(\tau)\}$. There are two key points
here.

The first and less important point is that this transformation included
the path integral representation of the equilibrium partition function,
resulting in the introduction of the set of imaginary time noises.
The dynamics associated with these noises is responsible for the joint
preparation of the total system (electrons and phonons), initialised
in the correct equilibrium state which includes quantum correlations
between the electronic and phonon sub-spaces rather than the two sub-spaces
being thermalized separately in the so-called partitioned approach.
This formulation of the equilibrium density matrix was first derived
in \cite{grabert1988quantum,moix2012equilibrium}, and then again
later in \cite{PhysRevLett.88.170407,PhysRevB.95.125124}, and used
by Tanimura\cite{tanimura2014reduced} to develop hierarchical equations
of motion for fermionic systems with an Ohmic spectral density for
the environment, and by Lane \textit{et al.}\cite{PhysRevB.101.224306}
to quantify deviations from expected asymptotic results which occur
as a result of system memory of the nonphysical partitioned state.
As a result of this formulation, the contact between electrons and
phonons in the molecular junction is neither partitioned nor approximate,
removing any spurious transient dynamics associated with the initial
mixing of artificially separated sub-spaces which would always otherwise
be present\cite{hilt2011hamiltonian}. In fact, in the approach we
have developed, memory of the initial preparation appears explicitly
in the form of the cross-time correlation function between the set
of real time noises $\{\eta_{A\alpha}(t)\}$ and imaginary time noises
$\{\overline{\mu}_{A\alpha}(\tau)\}$, which is equivalent to the
initial entanglement between electrons and phonons having a direct
impact on the subsequent system dynamics.

The second and more important point is that the introduction of these
noises causes the system dynamics to be non-Hermitian, a property
which manifests in three main features: the Hamiltonian is different
on the upper and lower horizontal branches of the contour; the forward
and backward propagators of the electronic density matrix in the Liouville
equation are no longer Hermitian conjugates of each other; and the
trace of the electronic density matrix is not preserved over the dynamics.
Dealing with this non-Hermicity isues requires the generalization
of the NEGF formalism to include the stochastic branch dependent time
evolution operators that we have presented here.

The first of these three features, that the Hamiltonian is now sensitive
to the upper and lower branches, supersedes the other two as it is
in some sense responsible for them. Since the dynamics over shared
periods of time on the branches no longer annul each other, we introduced
a third time argument into the NEGF, the observation time $\overline{t}$,
which is the right most time on the contour, as this controls the
extent of the full dynamics on the horizontal branches of the contour.
The definition of this three-time Green's function $G(z,z\'\vert\overline{t})$
in Eq. \eqref{eq:threetimeG} requires the introduction of a multiplicative
prefactor which is a function of the observation time alone, $\Pi(\overline{t})$,
which would otherwise appear in its equation of motion. Though an
expression for $\Pi(\overline{t})$ was found, Eq. \eqref{eq:Pi2},
its initial value when $\overline{t}=t_{0}$ was not immediately obvious,
and required the definition of a similar function $\overline{\Pi}(\overline{\tau})$,
this time a function of the preparation time $\overline{\tau}$ which
defines a subbranch within the vertical branch, $\tau\in(0,\overline{\tau})$.
This provided the initial value, $\Pi(t_{0})=\overline{\Pi}(\beta\hbar)$,
but the expression for $\overline{\Pi}(\overline{\tau})$ Eq. \eqref{eq:Pibar}
introduced new imaginary time self-energies $\overline{\Sigma}_{CC}(\tau,\tau\'\vert\overline{\tau})$
and $\overline{\Lambda}_{CC}(\tau,\tau\'\vert\overline{\tau})$ and,
more importantly, a second three-time \textit{thermal} NEGF $\overline{G}(\tau,\tau\'\vert\overline{\tau})$,
defined exclusively on the subbranch parameterized by the preparation
time.

Using the Kadanoff-Baym equations of motion for these three-time Green's
functions, we obtained a perturbative expansion Eq. \eqref{eq:explicitDyson}
in terms of the unravelling matrix which contains the noises and the
phonon-free Green's function for the molecular junction in the absence
of phonons. With a procedure to compute the three-time NEGFs in place,
we derived an expression for the non-equilibrium current response
to an external bias on the leads in Eq. \eqref{eq:current} which
involves the three-time NEGF, $\Pi(\overline{t})$, $\overline{\Pi}(\beta\hbar)$,
and a third new self-energy $\Upsilon_{CLC}(z,z\')$, which required
the application of a Girsanov transformation presented in Appendix
\ref{A:Girsanov} to remove a problematic infinity. Finally, expressions
for the self-energies in terms of energy integrals were derived, presented
in Appendix \ref{A:selfenergies}. Performing stochastic averages
over the realizations of the noises, the physical dynamics of the
system is then recovered, and this represents an exact, fully general,
and elegant framework for the inelastic dynamics of an electronic
open system coupled to a phonon environment. Of course, in practice
one would need to use a finite number of terms in the Born-like expansion
of the Green's functions leading to an approximate solution, and this
finite expansion would need to converge.

For clarity, we present here a condensed form of the procedure to
calculate the non-equilibrium current response to an external bias
through the leads using this method:
\begin{enumerate}
\item Compute quantities which are independent of the noises: components
of the phonon-free NEGF $G_{CC}^{0}(z,z\')$ from \cite{PhysRevB.91.125433},
$\overline{\Sigma}_{CC}^{>}(\tau,0\vert\overline{\tau})$ Eq. \eqref{eq:Sigmabar1}
and $\overline{\Sigma}_{CC}^{<}(0,\tau\vert\overline{\tau})$ Eq.
\eqref{eq:Sigmabar2}, $\overline{\Lambda}_{CC}(\tau,\tau\'\vert\overline{\tau})$
Eq. \eqref{eq:Lambdabar} and various components of $\Upsilon_{CLC}(z,z\')$
(Appendix \ref{A:selfenergies}).
\item Generate realizations of $\eta_{A\alpha}$, $\nu_{A\alpha}$ and $\overline{\mu}_{A\alpha}$.
For each realization:
\begin{enumerate}
\item For $\overline{\tau}\in(0,\beta\hbar):$
\begin{enumerate}
\item Use Born-like expansion Eq. \eqref{eq:explicitDyson} on the subbranch
of $\kappa^{M}$ up to $\overline{\tau}$ to calculate the three-time
thermal NEGF $\overline{G}_{CC}(\tau,\tau\'\vert\overline{\tau})$.
\item Calculate $\overline{\Pi}(\overline{\tau})$ Eq. \eqref{eq:Pibar}.
\end{enumerate}
\item Initialize $\Pi(t_{0})=\overline{\Pi}(\beta\hbar)$.
\item For $\overline{t}\in(t_{0},t_{max})$:
\begin{enumerate}
\item Use Born-like expansion Eq. \eqref{eq:explicitDyson} on the entire
contour $\kappa$ to calculate the three-time NEGF $G_{CC}(z,z\'\vert\overline{t})$.
\item Calculate $\Pi(\overline{t})$ Eq. \eqref{eq:Pi2}.
\item Evaluate $\int_{\kappa}dz_{1}dz_{2}\Upsilon_{CLC}(z_{2},z_{1})G_{CC}(z_{1},z_{2}\vert\overline{t})$.
\end{enumerate}
\end{enumerate}
\item Evaluate stochastic averages $\langle\overline{\Pi}(\beta\hbar)\rangle_{\xi\overline{\xi}}$
and $\langle\Pi(\overline{t})\int_{\kappa}dz_{1}dz_{2}\Upsilon_{CLC}(z_{2},z_{1})G_{CC}(z_{1},z_{2}\vert\overline{t})\rangle_{\xi\overline{\xi}}.$
\item Differentiate the latter with respect to $\overline{t}$.
\item Evaluate the current Eq. \eqref{eq:current}.
\end{enumerate}
The feasibility of a numerical scheme involving the three-time NEGF
depends on the size of the phonon environment, since there are three
noises per phonon, and the computational cost scales with the total
number of noises that must be generated, as well as the number of
noises involved in a single realization of the system dynamics. Generating
the noises themselves should not be a problem, as optimized schemes
for generating noises of this kind already exist\cite{matos2020efficient},
and should be easily extendable to the NEGF framework. It will also
require evaluation of the integrals in Appendix \ref{A:selfenergies}
for the self energies, which may in some cases involve the use of
Matsubara sums, the Pad$\acute{\textnormal{e}}$ approximation, or
an extension of the wide band approximation (WBA); this is left for
future work. Therefore, the first application of the three-time NEGF
is likely to be for a small molecular junction whose central region
is on the order of single atoms, with a minimal number of phonon modes.

We are currently working on a numerical implementation of this method
for a simple junction, as well as on its further development in which
the noises are integrated out analytically.

\section{Conclusion}

\label{sec:conclusion} Using the influence functional formalism,
the electronic density matrix was obtained by reducing the total density
matrix with respect to phonons for a molecular junction with electrons
coupled to a phonon environment in the central region, attached to
an arbitrary number of leads. Phonon degrees of freedom were fully
removed and replaced by complex coloured Gaussian noises via application
of a two-time HS transformation, leading to a stochastic Liouville
equation for the dynamics of the electronic density matrix. This prompted
the definition of a three-time stochastic NEGF $G(z,z\'\vert\overline{t})$
to account for the non-Hermicity of the dynamics, encapsulated by
the dependence on the third time $\overline{t}$ which is the rightmost
time on the Konstantinov-Perel' contour, or the \textit{observation
time}. Initialization of this three-time NEGF involved the definition
of a second three-time \textit{thermal} NEGF $\overline{G}(\tau,\tau\'\vert\overline{\tau})$,
defined only on a subbranch of the Matsubara branch up to the third
time $\overline{\tau}$, the \textit{preparation time}. A perturbative
expansion for these new NEGFs in terms of the noises and the regular
phonon-free Green's function was derived, as well as an expression
for the non-linear current response to an arbitrary external bias
applied to the leads. Physical quantities are recovered by integrating
over the functional distribution of the noises. This represents an
extension of the NEGF formalism to include stochastic dynamics, or
more generally any non-Hermitian dynamics that is sensitive to the
upper and lower horizontal branches.

\section*{Acknowledgements}

M.A.L. is supported by the EPSRC Centre for Doctoral Training in Cross-Disciplinary
Approaches to NonEquilibrium Systems (CANES, Grant No. EP/L015854/1).

\clearpage\newpage{}

\appendix

\section*{Appendices}

\section{Hubbard-Stratonovich and the Influence Functional}

\label{A:influence} By transforming into the normal mode coordinates
of the bath $q_{\lambda}$, the central region Hamiltonian becomes
diagonal with respect to the phonons (indicated by $\lambda$), 
\begin{gather}
\hatH_{C}=\hatH_{C}^{0}+\frac{1}{2}\sum_{\lambda}\left[p_{\lambda}^{2}+\omega_{\lambda}^{2}q_{\lambda}^{2}\right]-\sum_{\lambda}\hat{\sigma}_{\lambda}q_{\lambda}.\label{eq:HCmode}\\
q_{\lambda}=\sum_{A\alpha}\sqrt{m_{A\alpha}}e_{\lambda,A\alpha}u_{A\alpha},\label{eq:normalmodeDef}
\end{gather}
where $e_{\lambda}=\{e_{\lambda,A\alpha}\}$ are the eigenvectors
of the bath's dynamical matrix $D_{AA\'}^{\alpha\alpha\'}=\Lambda_{AA\'}^{\alpha\alpha\'}/\sqrt{m_{A\alpha}m_{A\'\alpha\'}}$
with associated eigenvalues $\omega_{\lambda}^{2}$.

The density matrix of the entire system, being a solution of the Liouville
equation, evolves in time via (in this Appendix, we use $t$ instead
of the observation time $\overline{t}$ to avoid cumbersome notations)
\begin{equation}
\rho_{tot}(t)=\hat{U}(t,t_{0})\rho_{tot}(t_{0})\hat{U}(t_{0},t)
\end{equation}
and therefore, can be thought of as a forward propagation along the
real time from $t_{0}$ to $\overline{t}$ by the regular time evolution
operator $\hat{U}(t,t_{0})$ which is not stochastic (this corresponds
to the upper branch $\kappa^{+}$) followed by the backward propagation
from $\overline{t}$ to $t_{0}$ via $\hat{U}(t_{0},t)=\hat{U}^{\dagger}(t,t_{0})$
(this corresponds to the lower branch $\kappa^{-}$). In the coordinate
representation the density matrix depends on the initial and final
coordinates of both the electrons and phonons, with all electronic
variables $x$ and phonon variables $q$ carrying a superscript $\pm$
which corresponds to the upper and lower horizontal branches, respectively.

The forward propagator of the total system from some initial time
$t_{1}$ up to a later time $t_{2}$ along the upper branch $\kappa^{+}$,
written in the coordinate representation with respect to electronic
coordinates $x$ and phonon coordinates $q$, can be expressed as
a path integral over electronic trajectories $x^{+}[s]$ and phonon
trajectories $q^{+}[s]$, 
\begin{align}
\langle x_{2},q_{2}\vert\hat{U}(t_{2},t_{1})\vert x_{1},q_{1}\rangle=\int_{x^{+}(t_{1})=x_{1}}^{x^{+}(t_{2})=x_{2}}\mathcal{D}\left[x^{+}(s)\right]\int_{q^{+}(t_{1})=q_{1}}^{q^{+}(t_{2})=q_{2}}\mathcal{D}\left[q^{+}(s)\right]e^{\frac{i}{\hbar}S_{tot}\left[x^{+}(s),q^{+}(s)\right]},
\end{align}
where $S_{tot}\left[x^{+}(s),q^{+}(s)\right]$ is the action of the
total system of electrons and phonons, 
\begin{align}
S_{tot}\left[x(s),q(s)\right]=S_{el}^{0}\left[x(s)\right]+S_{ph}\left[x(s),q[s]\right].
\end{align}
Here, $S_{el}^{0}\left[x(s)\right]$ is the classical action associated
with the isolated electronic subsystem of Eq. \eqref{eq:H01}, and
the action $S_{ph}\left[x(s),q[s]\right]$ contains all terms which
depend on the phonons from Eq. \eqref{eq:normalmodeDef}. Note that,
since phonons and electrons are coupled together, $S_{ph}$ must also
depend on the electronic trajectories.

Similarly, the backwards propagator of the total system is 
\begin{align}
\langle x_{1},q_{1}\vert\hat{U}(t_{1},t_{2})\vert x_{2},q_{2}\rangle=\int_{x^{-}(t_{2})=x_{2}}^{x^{-}(t_{1})=x_{1}}\mathcal{D}\left[x^{-}(s)\right]\int_{q^{-}(t_{2})=q_{2}}^{q^{-}(t_{1})=q_{1}}\mathcal{D}\left[q^{-}(s)\right]e^{-\frac{i}{\hbar}S_{tot}\left[x^{-}(s),q^{-}(s)\right]},
\end{align}
where the use of the $\pm$ superscripts clearly indicates which branch
$\kappa^{\pm}$ the evolution is on, and with respect to which time
ordering. For the backwards propagator, the limits of integration
are reversed to reflect the anti-chronological time ordering, with
the minus sign in the exponent coming from the time integral in the
action going from $t_{2}$ to $t_{1}$ with $t_{1}<t_{2}$.

The path integrals are performed with respect to both the open system
electronic trajectories and the phonon trajectories. However, the
integration over the environment (phonons) can be performed exactly
as the part of the Hamiltonian which involves phonons in Eq. \eqref{eq:normalmodeDef}
is that of a set of independent displaced harmonic oscillators. Consequently
the path integral is Gaussian and its result is well known\cite{feynman2000theory,feynman2010quantum,grabert1988quantum},
so that the propagators becomes path integrals over the open system
electronic trajectories only. For example, the forward propagator
is now given by, 
\begin{align}
\langle x_{2},q_{2}\vert\hat{U}(t_{2},t_{1})\vert x_{1},q_{1}\rangle=A^{+}\int_{x^{+}(t_{1})=x_{1}}^{x^{+}(t_{2})=x_{2}}\mathcal{D}\left[x^{+}(s)\right]e^{\frac{i}{\hbar}\left(S_{el}^{0}\left[x^{+}(s)\right]+S_{ph}\left[q_{2}^{+},q_{1}^{+},x^{+}(s)\right]\right)},\label{eq:Ucoord}
\end{align}
where and $q_{1}^{+}$ and $q_{2}^{+}$ are initial and final phonon
normal mode coordinates associated with the initial and final time
arguments in the time evolution operator, and 
\begin{align}
A^{\pm}=\prod_{\lambda}A_{\lambda}^{\pm}=\prod_{\lambda}\frac{1\mp i}{2\sqrt{\pi\hbar}}\sqrt{\frac{\omega_{\lambda}}{\sin\left(\omega_{\lambda}\left(t_{2}-t_{1}\right)\right)}}
\end{align}
is an oscillating amplitude which arises from a closed loop path integral
for each mode. The phonon part of the action, $S_{ph}\left[q_{2}^{+},q_{1}^{+},x^{+}(s)\right]$,
depends only on the initial and final values of the phonon variables,
but is still a functional of the electronic coordinates $x^{+}[s]$
along the trajectory.

The corresponding backwards propagator takes the same form, with the
replacement $+\rightarrow-$, a minus sign in the exponent, and the
limits of integration reversed as before.

The electronic only density matrix is obtained by taking the diagonal
element of the total density matrix with respect to the phonon coordinates
at the final time $\overline{t}$ and integrating over them
\begin{align}
\rho(x_{t},t;x_{0},t_{0}) & =\int dq_{t}\,\langle x_{t}q_{t}\vert\rhohat_{tot}(t)\vert x_{0}q_{t}\rangle\\
 & =\int dq_{t}dx_{1}dq_{1}dx_{2}dq_{2}\langle x_{t},q_{t}\vert\hat{U}(t,t_{0})\vert x_{2},q_{2}\rangle\langle x_{2},q_{2}\vert\rho_{tot}(t_{0})\vert x_{1},q_{1}\rangle\langle x_{1},q_{1}\vert\hat{U}(t_{0},t)\vert x_{0},q_{t}\rangle,\label{eq:rhohat}
\end{align}
that is, Eq. \eqref{eq:rhohat} is the reduced density matrix with
respect to the phonons.

Assuming that the total system was initially in thermal equilibrium,
\begin{align}
\rho_{tot}(t_{0})=\frac{1}{Z_{tot,0}}e^{-\beta\hat{\mathcal{H}}_{0}},
\end{align}
the part associated with the equilibrium density matrix at time $t_{0}$,
after integration over phonons, can also be expressed as an electronic
path integral, this time with a dummy imaginary time variable $\tau$,
\begin{align}
\langle x_{2},q_{\beta\hbar}^{M}\vert\rho_{tot}(t_{0})\vert x_{1},q_{0}^{M}\rangle=\frac{A^{M}}{Z_{tot,0}}\int_{x^{M}(0)=x_{1}}^{x^{M}(\beta\hbar)=x_{2}}\mathcal{D}\left[x_{M}(\tau)\right]e^{-\frac{1}{\hbar}\left(S_{el}^{E}\left[x^{M}(\tau)\right]+S_{ph}^{E}\left[q_{\beta\hbar}^{M},q_{0}^{M},x^{M}(\tau)\right]\right)},
\end{align}
where $S_{el}^{E}\left[x_{M}(\tau)\right]$ is the Euclidean action
where the electronic Hamiltonian Eq. \eqref{eq:H01} is used in place
of the Lagrangian in an integral over $\tau$ from 0 to $\beta\hbar$,
and $S_{ph}^{E}\left[q_{\beta\hbar}^{M},q_{0}^{M},x^{M}(\tau)\right]$
is the same but for the part of Eq. \eqref{eq:normalmodeDef} that
involves phonons and depends on the initial and final values of the
phonon coordinates at to $\tau=0$ and $\tau=\beta\hbar$, respectively.
Here, the label $M$ serves the same role as the $\pm$ labels used
previously, indicating that this imaginary time evolution can be thought
of as evolution along the vertical branch $\kappa^{M}$, and 
\begin{align}
A^{M}=\prod_{\lambda}A_{\lambda}^{M}=\prod_{\lambda}\sqrt{\frac{1}{2\pi\hbar}}\sqrt{\frac{\omega_{\lambda}}{\sinh\left(\beta\hbar\omega_{\lambda}\right)}}.
\end{align}

Eq. \eqref{eq:rhohat} now becomes 
\begin{align}
\rho(x_{t},t;x_{0},t_{0}) & =\frac{1}{Z_{0}}\int dx_{1}dx_{2}\int_{x^{+}(t_{0})=x_{2}}^{x^{+}(t)=x_{t}}\mathcal{D}\left[x^{+}(s)\right]\int_{x^{M}(0)=x_{1}}^{x^{M}(\beta\hbar)=x_{2}}\mathcal{D}\left[x^{M}(\tau)\right]\int_{x^{-}(t)=x_{0}}^{x^{-}(t_{0})=x_{1}}\mathcal{D}\left[x^{-}(s')\right]\nonumber \\
 & \quad\times\exp\left\{ \frac{i}{\hbar}\left(S_{el}^{0}\left[x^{+}(s)\right]-S_{el}^{0}\left[x^{-}(s')\right]+iS_{el}^{E}\left[x^{M}(\tau)\right]\right)\right\} \mathcal{F}[x^{+}(s),x^{-}(s'),x^{M}(\tau)]\label{eq:RDMF}
\end{align}
where 
\begin{align}
\mathcal{F}[x^{+}(s),x^{-}(s'),x^{M}(\tau)]=\frac{A^{+}A^{M}A^{-}}{Z_{ph}}\prod_{\lambda}\int dq_{\lambda}dq_{\lambda}^{+}dq_{\lambda}^{-}F_{\lambda}^{+}\left[q,q,x^{+}(s)\right]F_{\lambda}^{M}\left[q,q,x^{M}(\tau)\right]F_{\lambda}^{-}\left[q,q,x^{-}(s')\right]\label{eq:Ffunct}
\end{align}
is the influence functional which contains all the information about
the effect of the electron-phonon coupling on the electronic junction,
having already performed the path integration over the phonon trajectories.
Here, the total partition function $Z_{tot,0}$ has been split into
the partition function for the isolated phonon subsystem $Z_{ph}$,
\begin{align}
Z_{ph}=\frac{1}{2}\prod_{\lambda}\textnormal{cosech}\left(\frac{1}{2}\beta\hbar\omega_{\lambda}\right),
\end{align}
and the partition function for the remaining part $Z_{0}$, so that
$Z_{tot,0}=Z_{0}Z_{ph}$. The three functions $F^{\pm,M}$ come from
the path integrals on the three branches of the contour and are exponential
functions whose exponents are quadratic in the integration variables\cite{feynman2000theory,grabert1988quantum}.
Integration is performed for each phonon mode $\lambda$ separately,
and the quadratic structure of the exponent makes the integrals Gaussian
and therefore directly integrable, resulting in a single exponential
of a sum over modes, 
\begin{align}
\mathcal{F}[x^{+}(s),x^{-}(s'),x^{M}(\tau)]=\frac{A^{+}A^{M}A^{-}}{Z_{ph}}\exp\left\{ -\frac{1}{\hbar}\sum_{\lambda}\Phi_{\lambda}\left[x_{+}(s),x_{-}(s'),x_{M}(\tau)\right]\right\} ,\label{eq:fucntional-via-exp}
\end{align}
where $\Phi=\sum_{\lambda}\Phi_{\lambda}$ is the influence phase\cite{grabert1988quantum,PhysRevB.95.125124},
and the prefactor $A^{+}A^{M}A^{-}/Z_{ph}=1$, as can be shown by
a simple algebra, so that $\mathcal{F}[x^{+}(s),x^{-}(s'),x^{M}(\tau)]=\exp\left(-\Phi/\hbar\right)$.

This influence phase will depend on the coupling between the electronic
subsystem and the phonons. Recalling the electronic coupling operator
to the $A\alpha$ phonon $\hat{\sigma}_{A\alpha}$ from Eq. \eqref{eq:Hel-ph},
the operators associated with electronic states have been replaced
by classical trajectories in the path integrals, leading to the definition
of the branch dependent coupling functions $\sigma_{\lambda}^{\pm}(s)=\sigma_{\lambda}[x^{\pm}(s)]$
and $\sigma_{\lambda}^{M}(\tau)=\sigma_{\lambda}[x^{M}(\tau)]$ which
couple each phonon mode $\lambda$ to the electronic trajectory on
one of the three branches $x^{\pm}(s)$ or $x^{M}(\tau)$. After much
algebra \cite{PhysRevB.95.125124}, each mode of the influence phase
becomes 
\begin{align}
\Phi_{\lambda} & =\frac{1}{2}\int_{t_{0}}^{t}dt_{1}dt_{2}K_{\lambda}^{\textnormal{Re}}(t_{1}-t_{2})v_{\lambda}(t_{1})v_{\lambda}(t_{2})+2i\int_{t_{0}}^{t}dt_{1}dt_{2}\Theta(t_{1}-t_{2})K_{\lambda}^{\textnormal{Im}}(t_{1}-t_{2})v_{\lambda}(t_{1})w_{\lambda}(t_{2})\nonumber \\
 & \quad-i\int_{0}^{\beta\hbar}d\tau_{1}\int_{t_{0}}^{t}dt_{1}K_{\lambda}(t_{1}-i\tau_{1})\sigma_{\lambda}^{M}(\tau_{1})v_{\lambda}(t_{1})\nonumber \\
 & \quad-\frac{1}{2}\int_{0}^{\beta\hbar}d\tau_{1}d\tau_{2}\left[K_{\lambda}^{\textnormal{e}}(\tau_{1}-\tau_{2})-K_{\lambda}^{\textnormal{o}}(\lvert\tau_{1}-\tau_{2}\rvert)\right]\sigma_{\lambda}^{M}(\tau_{1})\sigma_{\lambda}^{M}(\tau_{2}),\label{eq:Phi_lambda}
\end{align}
where 
\begin{gather}
v_{\lambda}(t)={\sigma}_{\lambda}^{+}(t)-{\sigma}_{\lambda}^{-}(t)\qquad\textnormal{and}\qquad w_{\lambda}(t)=\frac{1}{2}\left[{\sigma}_{\lambda}^{+}(t)+{\sigma}_{\lambda}^{-}(t)\right]\\
K_{\lambda}^{\textnormal{Re}}(t)=\frac{1}{2\omega_{\lambda}}\cos(\omega_{\lambda}t)\coth\left(\frac{1}{2}\omega_{\lambda}\beta\hbar\right)\label{eq:KRe}\\
K_{\lambda}^{\textnormal{Im}}(t)=-\frac{1}{2\omega_{\lambda}}\sin(\omega_{\lambda}t)\label{eq:KIm}\\
K_{\lambda}(z)=\frac{1}{2\omega_{\lambda}}\frac{\cosh\left(\omega_{\lambda}\left(\frac{1}{2}\beta\hbar-iz\right)\right)}{\sinh\left(\frac{1}{2}\omega_{\lambda}\beta\hbar\right)}\label{eq:Klambda}\\
K_{\lambda}^{\textnormal{e}}(\tau)=\frac{1}{2\omega_{\lambda}}\coth\left(\frac{1}{2}\omega_{\lambda}\beta\hbar\right)\cosh(\omega_{\lambda}\tau)\label{eq:Ke}\\
K_{\lambda}^{\textnormal{o}}(\tau)=\frac{1}{2\omega_{\lambda}}\sinh(\omega_{\lambda}\tau).\label{eq:Kodd_lambda}
\end{gather}
Notably, these kernels satisfy the following relationships: that $K_{\lambda}(t)=K_{\lambda}^{\textnormal{Re}}(t)+iK_{\lambda}^{\textnormal{Im}}(t)$,
and $K_{\lambda}(i\tau)=K_{\lambda}^{\textnormal{e}}(\tau)+K_{\lambda}^{\textnormal{o}}(\tau)$,
with $K_{\lambda}^{\textnormal{e}}$ and $K_{\lambda}^{\textnormal{o}}$
being even and odd functions, respectively. The first two terms in
Eq. \eqref{eq:Phi_lambda} are the regular integrals which appear
in the Feynman-Vernon influence functional\cite{feynman2000theory},
while the remaining terms arise from considering the total system
to be in thermal equilibrium at $t_{0}$ rather than artificially
partitioned\cite{PhysRevB.95.125124}.

Returning to the site representation for the phonons with the joint
index $a=(A\alpha)$, the total influence phase in Eq. (\ref{eq:fucntional-via-exp})
is now 
\begin{align}
-\frac{1}{\hbar}\Phi & =-\frac{1}{\hbar}\sum_{aa'}\Phi_{aa'}=-\frac{1}{2}\left[\int_{t_{0}}^{t}dt_{1}\int_{t_{0}}^{t}dt_{2}\textnormal{v}_{a}^{T}(t_{1})\textnormal{\textbf{C}}_{aa\'}(t_{1}-t_{2})\textnormal{v}_{a\'}(t_{2})\right.\nonumber \\
 & \quad\left.+2\int_{0}^{\beta\hbar}d\tau_{1}\int_{t_{0}}^{t}dt_{1}\overline{\textnormal{v}}_{a}^{T}(\tau_{1})\textnormal{\textbf{C}}_{aa\'}^{\times}(t_{1},\tau_{1})\textnormal{v}_{a'}(t_{1})+\int_{0}^{\beta\hbar}d\tau_{1}\int_{0}^{\beta\hbar}d\tau_{2}\overline{\textnormal{v}}_{a}^{T}(\tau_{1})\mathbf{\overline{C}}_{aa\'}(\tau_{1}-\tau_{2})\overline{\textnormal{v}}_{a\'}(\tau_{2})\right]\label{eq:Phi_coord}
\end{align}
where 
\begin{gather}
\textnormal{v}_{a}(t)=\begin{pmatrix}v_{a}(t)/\hbar\\
0\\
w_{a}(t)\\
0
\end{pmatrix}\qquad\textnormal{and}\qquad\overline{\textnormal{v}}_{a}(\tau)=\begin{pmatrix}i\sigma_{a}^{M}(\tau)/\hbar\\
0
\end{pmatrix}\\
\textnormal{\textbf{C}}_{aa\'}(t)=\begin{pmatrix}\hbar K_{aa\'}^{\textnormal{Re}}(t) & 0 & 2i\Theta(t)K_{aa\'}^{\textnormal{Im}}(t) & 0\\
0 & 0 & 0 & 0\\
2i\Theta(-t)K_{aa\'}^{\textnormal{Im}}(-t) & 0 & 0 & 0\\
0 & 0 & 0 & 0
\end{pmatrix}\label{eq:R}\\
\textnormal{\textbf{C}}_{aa\'}^{\times}(t,\tau)=\begin{pmatrix}-\hbar K_{aa\'}(t-i\tau) & 0 & 0 & 0\\
0 & 0 & 0 & 0
\end{pmatrix}\label{eq:x}\\
\mathbf{\overline{C}}(\tau)=\begin{pmatrix}\hbar\left[K_{aa\'}^{\textnormal{e}}(\tau)-K_{aa\'}^{\textnormal{o}}(\lvert\tau\rvert)\right] & 0\\
0 & 0
\end{pmatrix}\label{eq:Rbar}
\end{gather}
so that the total influence phase is now $\Phi=\sum_{aa\'}\Phi_{aa\'}$,
having used transformations of the form 
\begin{align}
K_{aa\'}=\frac{1}{\sqrt{m_{a}m_{a\'}}}\sum_{\lambda}e_{\lambda a}e_{\lambda a\'}K_{\lambda}.\label{eq:normalmode}
\end{align}
As before, the $e_{\lambda a}$ are the elements of the eigenvectors
of the bath's dynamical matrix.

The equations above have been written specifically in the form most
suitable for the Hubbard-Stratonovich transformation\cite{PhysRevLett.88.170407,PhysRevLett.3.77}
with complex multivariate Gaussian noises\cite{stockburger2004simulating}
that is applied with respect to the real and imaginary times\cite{PhysRevB.95.125124}.
This transformation maps the bi-linear exponent in Eq. \eqref{eq:Phi_coord}
(the total phase in Eq. \eqref{eq:fucntional-via-exp}) onto a stochastic
exponent which is linear in the noises, at the expense of then averaging
over all realizations of those noises, 
\begin{align}
\exp\left\{ -\frac{1}{\hbar}\sum_{aa\'}\Phi_{aa\'}\right\}  & =\Bigg\langle\exp\left\{ i\sum_{a}\left[\int_{t_{0}}^{t}dt_{1}\xi_{a}^{T}(t_{1})\textnormal{v}_{a}(t_{1})+\int_{0}^{\beta\hbar}d\tau_{1}\overline{\xi}_{a}^{T}(\tau_{1})\overline{\textnormal{v}}_{a}(\tau_{1})\right]\right\} \Bigg\rangle_{\xi\overline{\xi}},\label{eq:HS}
\end{align}
where 
\begin{align}
\xi_{a}(t)=\begin{pmatrix}\eta_{a}(t)\\
\eta_{a}^{*}(t)\\
\nu_{a}(t)\\
\nu_{a}^{*}(t)
\end{pmatrix}\qquad\textnormal{and}\qquad\overline{\xi}_{a}(\tau)=\begin{pmatrix}\overline{\mu}_{a}(\tau)\\
\overline{\mu}_{a}^{*}(\tau)
\end{pmatrix}\label{eq:xi}
\end{align}
are vectors of the noises and their complex conjugates, and $\langle\ldots\rangle_{\xi\overline{\xi}}$
indicates the average over $\xi$ and $\overline{\xi}$. It is worth
emphasising the structure of the noise vectors $\xi_{a}$ and $\overline{\xi}_{a}$:
they contain complex conjugate pairs of the noises. This gives meaning
to the matrices in Eqs. \eqref{eq:R}-\eqref{eq:Rbar}, which up until
now have simply been algebraic, when in fact they are the precision
matrices of the distribution functional of the noises. For example,
for $\textnormal{R}_{aa\'}$: the 11 component is the correlation
function of $\eta_{a}$ and $\eta_{a\'}$; the 12 component is the
correlation function of $\eta_{a}$ and $\eta_{a\'}^{*}$; the 13
component is the correlation function of $\eta_{a}$ and $\nu_{a\'}$,
and so on.

The noises therefore have the following correlation functions, 
\begin{gather}
\langle\eta_{a}(t)\eta_{a\'}(t\')\rangle_{\xi\overline{\xi}}=\hbar K_{aa\'}^{\textnormal{Re}}(t-t\')\label{eq:etaeta}\\
\langle\eta_{a}(t)\nu_{a\'}(t\')\rangle_{\xi\overline{\xi}}=2i\Theta(t-t\')K_{aa\'}^{\textnormal{Im}}(t-t\')\label{eq:etanu}\\
\langle\eta_{a}(t)\overline{\mu}_{a\'}(\tau)\rangle_{\xi\overline{\xi}}=-\hbar K_{aa\'}(t-i\tau)\label{eq:etamu}\\
\langle\overline{\mu}_{a}(\tau)\overline{\mu}_{a\'}(\tau\')\rangle_{\xi\overline{\xi}}=\hbar\left[K_{aa\'}^{\textnormal{e}}(\tau_{1}-\tau_{2})-K_{aa\'}^{\textnormal{o}}(\lvert\tau_{1}-\tau_{2}\rvert)\right],\label{eq:mumu}
\end{gather}
with all other correlations not shown being zero. The matrix elements
$\mathbf{C}_{aa\'},$ $\mathbf{C}_{aa\'}^{\times}$ and $\mathbf{\overline{C}}_{aa\'}$
of Eqs. \eqref{eq:R}-\eqref{eq:Rbar} are therefore identified as
the correlations between the $a$ and $a\'$ noises, which appear
as elements of a $2\times2$ block in the partitioned covariance matrix
between the noises,
\begin{equation}
\boldsymbol{\Sigma}_{aa\'}=\left(\begin{array}{cc}
\mathbf{C}_{aa\'} & \mathbf{C}_{aa\'}^{\times}\\
\mathbf{C}_{aa\'}^{\times T} & \mathbf{\overline{C}}_{aa\'}
\end{array}\right),\label{eq:covarianceMatrix}
\end{equation}
having been partitioned with respect to the real-time vector noises
$\xi_{a}(t)$ and the imaginary time vector noises $\overline{\xi}_{a}(\tau)$.
The full covariance matrix for all the $a$ and $a\'$ is thus $\boldsymbol{\Sigma}=\left(\boldsymbol{\Sigma}_{aa\'}\right)$.

It is worth emphasising the equality between Eq. \eqref{eq:HS} and
Eq. \eqref{eq:Phi_coord}: this is not an approximation. Rather, the
noises have been introduced in a mathematically exact way and their
properties rigorously derived from the theory, with the average of
Eq. \eqref{eq:HS} over the Gaussian distribution functional of the
noises being formally equivalent to Eq. \eqref{eq:Phi_coord}. Reincorporating
the influence functional back into the path integrals by inserting
Eqs. \eqref{eq:HS} and \eqref{eq:fucntional-via-exp} into Eq. \eqref{eq:RDMF},
we obtain 
\begin{align}
\widetilde{\rho}(x_{t},t;x_{0},t_{0}) & =\frac{1}{Z_{0}}\int dx_{1}dx_{2}\int_{x^{+}(t_{0})=x_{2}}^{x^{+}(t)=x_{t}}\mathcal{D}\left[x^{+}(s)\right]\int_{x^{M}(0)=x_{1}}^{x^{M}(\beta\hbar)=x_{2}}\mathcal{D}\left[x^{M}(\tau)\right]\int_{x^{-}(t)=x_{0}}^{x^{-}(t_{0})=x_{1}}\mathcal{D}\left[x^{-}(s')\right]\nonumber \\
 & \quad\times\exp\left\{ \frac{i}{\hbar}\left(S^{+}\left[x^{+}(s)\right]-S^{-}\left[x^{-}(s')\right]+i\overline{S}\left[x^{M}(\tau)\right]\right)\right\} 
\end{align}
for the electronic density matrix, where the three actions $S^{\pm}$
and $\overline{S}$ are now stochastic - hence the tilde over the
$\rho$ to indicate that it is a non-physical stochastic quantity
- and correspond to stochastic potentials in the Lagrangians, 
\begin{gather}
S^{\pm}[x^{\pm}(s)]=\int_{t_{0}}^{t}ds\left(L_{el}^{0}[x^{\pm}(s)]+\sum_{A\alpha}\left[\eta_{A\alpha}(s)\pm\frac{\hbar}{2}\nu_{A\alpha}(s)\right]\sigma_{A\alpha}^{\pm}(s)\right)\label{eq:L1}\\
\overline{S}[x^{M}(\tau)]=\int_{0}^{\beta\hbar}d\tau\left(L_{el}^{0}[x^{M}(\tau)]+\sum_{A\alpha}\overline{\mu}_{A\alpha}(\tau)\sigma_{A\alpha}^{M}(\tau)\right),\label{eq:L2}
\end{gather}
where $L_{el}^{0}$ is the purely electronic (phonon-free) Lagrangian
associated with the Hamiltonian in Eq. \eqref{eq:H01}. The new stochastic
Lagrangians which are the full integrands of Eqs. \eqref{eq:L1} and
\eqref{eq:L2} have the corresponding stochastic Hamiltonians that
are precisely those of Eq. \eqref{eq:Hkappa}.

Returning to the propagators associated with each of the path integrals,
and using the same coordinates as in Eq. \eqref{eq:rhohat}, we obtain
\begin{gather}
\langle x_{t}\lvert\hatU^{+}(t,t_{0})\rvert x_{2}\rangle=\int_{x^{+}(t_{0})=x_{2}}^{x^{+}(t)=x_{t}}\mathcal{D}\left[x^{+}(s)\right]e^{\frac{i}{\hbar}S^{+}\left[x^{+}(s)\right]}\\
\langle x_{2}\lvert\widetilde{\rho}_{0}\rvert x_{1}\rangle=\int_{x^{M}(0)=x_{1}}^{x^{M}(\beta\hbar)=x_{2}}\mathcal{D}\left[x^{M}(\tau)\right]e^{-\frac{1}{\hbar}\overline{S}\left[x^{M}(\tau)\right]}\\
\langle x_{1}\lvert\hatU^{-}(t_{0},t)\rvert x_{0}\rangle=\int_{x^{-}(t)=x_{0}}^{x^{-}(t_{0})=x_{1}}\mathcal{D}\left[x^{-}(s)\right]e^{-\frac{i}{\hbar}S^{-}\left[x^{-}(s)\right]},
\end{gather}
where $\mathcal{U}^{\pm}$ and $\overline{\mathcal{U}}(\beta\hbar,0)=\widetilde{\rho}_{0}$
are the stochastic time evolution operators on the upper and lower
horizontal branches and the vertical branch, respectively, which use
these new stochastic Hamiltonians. This brings us to the Liouville
equation of Eq. \eqref{eq:Liouville_noises} and completes the transformation.

\section{Contour Integrals}

\label{A:langreth} The regular Langreth rules cannot be applied when
expanding the contour integrals in Eqs. \eqref{eq:explicitDyson}
and \eqref{eq:current} since the unravelled Hamiltonian of Eqs. \eqref{eq:hkappa}-\eqref{eq:Hkappa}
is sensitive to all three of the branches $\kappa^{\pm}$ and $\kappa^{M}$.
Instead, the generalized Langreth rules\cite{PhysRevB.101.165408}
must be applied to consider all possible combinations of complex times
where real times on the upper and lower branches are treated separately
due to the different values of the unraveling matrix $W_{CC}$, and
always ordered with respect to the contour time ordering $\hat{\mathcal{T}}_{\kappa}$.

Adopting the following convention for the integration of three-time
quantities such as with the unravelling matrix $W_{CC}(z_{1},z_{2})$
in the expansion for $G_{CC}$ (Eq. \eqref{eq:explicitDyson}), 
\begin{equation}
\int_{\kappa}dz_{1}dz_{2}A(z,z_{1}\vert\overline{t})W(z_{1},z_{2})B(z_{2},z\'\vert\overline{t})=\begin{cases}
\left(A\bullet w\bullet B\right)(z,z,\'\vert\overline{t}) & t\in\kappa^{+}\\
\left(A\circ w\circ B\right)(z,z\'\vert\overline{t}) & t\in\kappa^{-}\\
\left(A\star\overline{w}\star B\right)(z,z\'\vert\overline{t}) & \tau\in\kappa^{M},
\end{cases}
\end{equation}
with
\begin{gather}
\left(A\bullet w\bullet B\right)(z,z,\'\vert\overline{t})\equiv\int_{t_{0}^{+}}^{\overline{t}}dt\,A(z,t\vert\overline{t})w(t)B(t,z\'\vert\overline{t}),\quad t\in\kappa^{+}\label{eq:bullet}\\
\left(A\circ w\circ B\right)(z,z\'\vert\overline{t})\equiv-\int_{t_{0}^{-}}^{\overline{t}}dt\,A(z,t\vert\overline{t})w(t)B(t,z\'\vert\overline{t}),\quad t\in\kappa^{-}\label{eq:circ}\\
\left(A\star\overline{w}\star B\right)(z,z\'\vert\overline{t})\equiv-i\int_{0}^{\beta\hbar}d\tau\,A(z,\tau\vert\overline{t})\overline{w}(\tau)B(\tau,z\'\vert\overline{t}),\quad\tau\in\kappa^{M},\label{eq:star}
\end{gather}
where we have exploited the $\delta-$functions in the definition
of $W_{CC}$ in Eq. \eqref{eq:W_CC-2-times}, we can write down integrals
of the product of functions defined on the contour explicitly. Using
the fact that only the unravelling matrix depends on the particular
horizontal branch, with the phonon-free Green's function $G^{0}$
being the same on both branches, the perturbative expansion for the
$+-$ component of the three-time NEGF in the central region is readily
expanded out, e.g., to first order, as 
\begin{align}
G_{CC}^{+-}(t,t\'\vert\overline{t}) & =G_{CC}^{0<}(t,t\')+\left[\left(G_{CC}^{0r}+G_{CC}^{0<}\right)\bullet w_{CC}\bullet G_{CC}^{0<}+G_{CC}^{0<}\circ w_{CC}\circ\left(G_{CC}^{0a}-G_{CC}^{0<}\right)\right.\nonumber \\
 & \quad\left.+G_{CC}^{0\rightGF}\star\overline{w}_{CC}\star G_{CC}^{0\leftGF}\right](t,t\')+\dots,\label{eq:G_expansion_1st_order}
\end{align}
where the dependence on the observation time $\overline{t}$ on the
right hand side is hidden in the definitions of the integrals for
$\bullet$ and $\circ$. The expressions for the components of the
phonon-free NEGF (e.g., in the wide band approximation\cite{stefanucci_vanleeuwen_2013}
they are given in Ref. \cite{PhysRevB.91.125433}) can then be used
to compute the three-time Green's function as required. Higher order
terms can be written similarly by a repetitive application of the
generalised Langreth rules \cite{PhysRevB.101.165408}.

The contour integral which appears in the current Eq. \eqref{eq:current}
is less straightforward since it involves integrating over both the
inner and outer times of $\Upsilon_{CLC}(z_{2},z_{1})G_{CC}(z_{1},z_{2}\vert\overline{t})$.
This calculation is easier to perform by representing each $z$ integral
as a sum over either of the three branches, leading to 9 terms. Since
one of the times in each of the two isolated lead Green's function
$g_{LL}^{0}$ in the definition of $\Upsilon_{CLC}(z_{2},z_{1})$
is fixed just before or after the observation time, Eq. \eqref{eq:Y-self-energy-for-current},
it is possible to indicate explicitly as superscripts in the self-energy
the particular components of the isolated leads Green's function,
\begin{equation}
\Upsilon_{CLC}^{\gamma\gamma'}(z_{2},z_{1})=h_{CL}g_{LL}^{0\gamma}(z_{2},\overline{t}^{-})g_{LL}^{0\gamma'}(\overline{t}^{+},z_{1})h_{LC}\,,
\end{equation}
where $\gamma,\gamma'=<,>,\rightGF,\leftGF$. Then, by considering
each possible combination of the branches, one obtains 
\begin{equation}
\begin{aligned}\int_{\kappa}dz_{1}dz_{2}\,\text{tr}\left[\Upsilon_{CLC}(z_{2},z_{1})G_{CC}(z_{1},z_{2}\vert\overline{t})\right] & =\int_{t_{0}}^{\overline{t}}dt_{1}dt_{2}\,\text{tr}\left[\Upsilon_{CLC}^{<>}(t_{2},t_{1})G_{CC}^{++}(t_{1},t_{2}\vert\overline{t})-\Upsilon_{CLC}^{<<}(t_{2},t_{1})G_{CC}^{-+}(t_{1},t_{2}\vert\overline{t})\right.\\
 & \left.\qquad-\Upsilon_{CLC}^{>>}(t_{2},t_{1})G_{CC}^{+-}(t_{1},t_{2}\vert\overline{t})+\Upsilon_{CLC}^{><}(t_{2},t_{1})G_{CC}^{--}(t_{1},t_{2}\vert\overline{t})\right]\\
 & -i\int_{t_{0}}^{\overline{t}}dt\int_{0}^{\beta\hbar}d\tau\,\text{tr}\left[\Upsilon_{CLC}^{<\rightGF}(t,\tau)G_{CC}^{M+}(\tau,t\vert\overline{t})-\Upsilon_{CLC}^{>\rightGF}(t,\tau)G_{CC}^{M-}(\tau,t\vert\overline{t})\right.\\
 & \qquad\left.+\Upsilon_{CLC}^{\leftGF>}(\tau,t)G_{CC}^{+M}(t,\tau\vert\overline{t})-\Upsilon_{CLC}^{\leftGF<}(\tau,t)G_{CC}^{-M}(t,\tau\vert\overline{t})\right]\\
 & -\int_{0}^{\beta\hbar}d\tau_{1}d\tau_{2}\,\text{tr}\left[\Upsilon_{CLC}^{\leftGF\ \rightGF}(\tau_{2},\tau_{1})G_{CC}^{MM}(\tau_{1},\tau_{2}\vert\overline{t})\right].
\end{aligned}
\label{eq:trYG}
\end{equation}
The components of $\Upsilon_{CLC}$ are given in Appendix \ref{A:selfenergies}.
This expression shows that alongside the component $G_{CC}^{+-}$
of the central region Green's function obtained in Eq. (\ref{eq:G_expansion_1st_order}),
one also needs similar expressions for other components, such as $G_{CC}^{\pm\pm}$,
$G_{CC}^{\pm M}$, $G_{CC}^{M\pm}$ and $G_{CC}^{MM}$. These can
be obtained using rules presented in \cite{PhysRevB.101.165408} in
a straightforward manner.

\section{Girsanov Transformation of the Noise Measure}

\label{A:Girsanov} Taking the average of $\Pi(\overline{t})$ in
Eq. \eqref{eq:lead-L-population}, the average can be treated as either
a statistical average over realizations of the noises, or as the functional
integral over their distribution $\mathcal{M}$, 
\begin{align}
\langle\Pi(\overline{t})\rangle_{\xi\overline{\xi}}=\int_{\xi\overline{\xi}}\mathcal{D}[\xi(\overline{t})]\mathcal{D}[\overline{\xi}(\tau)]\,\mathcal{M}[\xi(\overline{t}),\overline{\xi}(\tau)]\,\Pi(\overline{t}),\label{eq:Piaverage}
\end{align}
where $\xi_{a}$ and $\overline{\xi}_{a}$ are the vector noises of
the $a=(A\alpha)$ displacement from Eq. \eqref{eq:xi}, and $\xi=\{\xi_{a}\}$
and $\overline{\xi}=\{\overline{\xi}_{a}\}$ are the sets of noises
over all the displacements. The noise measure takes the form 
\begin{align}
\mathcal{M}[\xi(\overline{t}),\overline{\xi}(\tau)] & =\mathbb{M}\exp\left\{ -\frac{1}{2}\sum_{aa\'}\left(\begin{array}{c}
\xi_{a}\\
\overline{\xi}_{a}
\end{array}\right)^{T}\left(\begin{array}{cc}
\mathbf{A}_{aa\'} & \mathbf{B}_{aa\'}\\
\mathbf{B}_{aa\'}^{T} & \mathbf{D}_{aa\'}
\end{array}\right)\left(\begin{array}{c}
\xi_{a\'}\\
\overline{\xi}_{a\'}
\end{array}\right)\right\} \\
 & =\mathbb{M}\exp\left\{ -\frac{1}{2}\sum_{aa\'}\left[\int_{t_{0}}^{\overline{t}}dt_{1}\int_{t_{0}}^{\overline{t}}dt_{2}\ \xi_{a}^{T}(t_{1})\textnormal{\textbf{A}}_{aa\'}(t_{1}-t_{2})\xi_{a\'}(t_{2})\right.\right. \nonumber \\
 & \quad\left.\left.+2\int_{t_{0}}^{\overline{t}}dt_{1}\int_{0}^{\tau}d\tau_{1}\ \xi_{a}^{T}(t_{1})\textnormal{\textbf{B}}_{aa\'}(t_{1},\tau_{1})\overline{\xi}_{a\'}(\tau_{1})\right.\right. \nonumber \\
 & \quad\left.\left.+\int_{0}^{\tau}d\tau_{1}\int_{0}^{\tau}d\tau_{2}\ \overline{\xi}_{a}^{T}(\tau_{1})\textnormal{\textbf{D}}_{aa\'}(\tau_{1}-\tau_{2})\overline{\xi}(\tau_{2})\right]\right\} ,
\end{align}
where $\mathbb{M}$ is the Gaussian normalization factor, and the
matrix
\begin{equation}
\mathbf{P}_{aa\'}(z,z\')=\left(\begin{array}{cc}
\mathbf{A}_{aa\'}(t,t\') & \mathbf{B}_{aa\'}(t,\tau)\\
\mathbf{B}_{aa\'}^{T}(\tau,t) & \mathbf{D}_{aa\'}(\tau,\tau\')
\end{array}\right)
\end{equation}
is the precision matrix of the distribution; specifically, it is the
particular component of the full precision matrix $\mathbf{P}=\left(\textnormal{\textbf{P}}_{aa\'}\right)$,
between the $a$ and $a\'$ noises. This matrix shares the same structure
as the covariance matrix $\boldsymbol{\Sigma}_{aa\'}$ (Eq. \eqref{eq:covarianceMatrix}),
in which the blocks are matrices themselves which correspond to the
different pairs of times $(t,t\')$, $(t,\tau)$ and $(\tau,\tau\')$.
The full precision matrix $\mathbf{P}$ therefore has components for
both the sets of noises and pairs of times. For example, $\left(\mathbf{P}_{11}\right)_{aa'}=P_{1a,1a'}$
is the precision matrix between noises $\xi_{a}(t)$ and $\xi_{a'}(t')$
for both times being real (which is still a $4\times4$ matrix with
respect to the components of the noises), $\left(\mathbf{P}_{12}\right)_{aa'}=P_{1a,2a'}$
is the rectangular $4\times2$ matrix block for the noises $\xi_{a}(t)$
and $\overline{\xi}_{a'}(\text{\ensuremath{\tau}})$ corresponding
to one time real and one imaginary, and, finally, $\left(\mathbf{P}_{22}\right)_{aa'}=P_{2a,2a'}$
is the $2\times2$ block corresponding to the noises $\overline{\xi}_{a}(\tau)$
and $\overline{\xi}_{a'}(\tau')$ for both imaginary times. The full
matrix $\mathbf{P}$ is directly related to the covariance matrix
$\boldsymbol{\Sigma}$ of Eq. \eqref{eq:covarianceMatrix} by its
inverse, $\textnormal{\textbf{P}}_{aa\'}=\left(\boldsymbol{\Sigma}^{-1}\right)_{aa\'}$.
Thus the block matrices $\mathbf{C}_{aa\'}$, $\mathbf{C}_{aa\'}^{\times}$
and $\mathbf{\overline{C}}_{aa\'}$ which contain the correlation
functions between the noises are related to the precision matrix by
$\mathbf{C}_{aa\'}=(\textnormal{\textbf{P}}^{-1})_{1a,1a'}$, $\mathbf{C}_{aa\'}^{\times}=(\textnormal{\textbf{P}}^{-1})_{1a,2a'}$
and $\mathbf{\overline{C}}_{aa\'}=(\textnormal{\textbf{P}}^{-1})_{2a,2a'}$.

To simplify the notation, we shall henceforth ignore the complex conjugate
components since all correlation functions which involve complex conjugate
noises are equal to zero. The vector $\xi_{a}(t)$ then has only two
components $\eta_{a}$ and $\nu_{a}$, and $\overline{\xi}_{a}(\tau)$
has only one component $\overline{\mu}_{a}$. Next, we shall discretise
both real and imaginary times and introduce three blocks for the noises:
$\boldsymbol{\eta}=\left(\eta_{a}(t)\right)$, $\boldsymbol{\nu}=\left(\nu_{a}(t)\right)$,
and $\overline{\boldsymbol{\mu}}=\left(\overline{\mu}_{a}(\tau)\right)$,
with the full noise vector $\boldsymbol{\chi}=\left(\begin{array}{ccc}
\boldsymbol{\eta} & \boldsymbol{\nu} & \boldsymbol{\overline{\mu}}\end{array}\right)^{T}$. The noise measure can then be compactly written as 
\begin{equation}
\mathcal{M}[\xi(\overline{t}),\overline{\xi}(\tau)]=\mathbb{M}\exp\left\{ -\frac{1}{2}\boldsymbol{\chi}^{T}\mathbf{P}\boldsymbol{\chi}\right\} .
\end{equation}

Given that $\Pi(\overline{t})$ of Eq. \eqref{eq:Pi2} is an exponential
which is linear in the $\nu_{a}(t)$ noises (which are the second
component of $\boldsymbol{\chi}$), it can be written as
\begin{align}
\Pi(\overline{t})=\Pi_{0}\exp\left(\boldsymbol{\chi}^{T}\mathbf{L}\right),
\end{align}
where we have introduced a 3-component vector $\mathbf{L}=\left(\begin{array}{ccc}
\mathbf{0} & \boldsymbol{\gamma} & \mathbf{0}\end{array}\right)^{T}$, with 
\begin{equation}
\boldsymbol{\gamma}=\left(\gamma_{a}(t)\right)=\left(\hbar\,\text{tr}\left[\mathcal{V}_{CC}^{a}G_{CC}^{+-}(t,t\vert t)\right]\right).
\end{equation}
The Girsanov transformation \cite{PhysRevB.101.224306,girsanov1960transforming}
can then be applied such that the average taken over the transformed
measure $\mathcal{M}\'$ with respect to the transformed noises $\xi\',\overline{\xi}\'$
is analytically equivalent to the original average in Eq. \eqref{eq:Piaverage}
taken over the original measure $\mathcal{M}$ with respect to the
un-transformed noises $\xi,\,\overline{\xi}$: 
\begin{align}
\mathcal{M}[\xi(\overline{t}),\overline{\xi}(\tau)]\,\Pi[\xi(\overline{t})]=\Pi_{0}\,\mathcal{M}\'[\xi\'(\overline{t}),\overline{\xi}\'(\tau)]\,.\label{eq:Mtransform}
\end{align}

\noindent This transformation is achieved by completing the square
in the total exponent of the right hand side of Eq. \eqref{eq:Mtransform}
to introduce a new vector of noises, 
\begin{equation}
\boldsymbol{\chi}'=\boldsymbol{\chi}-\mathbf{P}^{-1}\mathbf{L}=\boldsymbol{\chi}-\boldsymbol{\Sigma}\mathbf{L},\label{eq:new-noises-vect}
\end{equation}
which transforms the quadratic form in the exponential $-\frac{1}{2}\boldsymbol{\chi}^{T}\mathbf{P}\boldsymbol{\chi}+\boldsymbol{\chi}^{T}\mathbf{L}$
into $\phi-\frac{1}{2}\left(\boldsymbol{\chi}'\right)^{T}\mathbf{P}\boldsymbol{\chi}'$,
where the free (noise independent) term,
\begin{equation}
\phi=\frac{1}{2}\mathbf{L}^{T}\mathbf{P}^{-1}\mathbf{L}=\frac{1}{2}\mathbf{L}^{T}\boldsymbol{\Sigma}\mathbf{L}
\end{equation}
\begin{equation}
=\frac{1}{2}\left(\begin{array}{ccc}
\mathbf{0} & \boldsymbol{\gamma} & \mathbf{0}\end{array}\right)\left(\begin{array}{ccc}
\mathbf{C}^{\eta\eta} & \mathbf{C}^{\eta\nu} & \mathbf{C}^{\times}\\
\mathbf{C}^{\nu\eta} & \mathbf{0} & \mathbf{0}\\
\mathbf{C}^{\times T} & \mathbf{0} & \mathbf{\overline{C}}
\end{array}\right)\left(\begin{array}{c}
\mathbf{0}\\
\boldsymbol{\gamma}\\
\mathbf{0}
\end{array}\right)=0\,,
\end{equation}
is equal to zero since the correlation functions between the $\nu$
noises, $\textnormal{C}_{aa'}^{\nu\nu}(t_{1},t_{2})=\langle\nu_{a}(t_{1})\nu_{a\'}(t_{2})\rangle$,
is itself equal to zero.

This is a very important point that the free term $\phi=0$; without
it, the application of the Girsanov transformation would introduce
an additional time dependence to Eq. \eqref{eq:NumberL} when differentiating
to obtain the current.

It is clear from Eq. \eqref{eq:new-noises-vect} that only the $\eta$
noises are affected by the transformation:
\begin{equation}
\left(\begin{array}{c}
\boldsymbol{\eta}^{\prime}\\
\boldsymbol{\nu}^{\prime}\\
\overline{\boldsymbol{\mu}}^{\prime}
\end{array}\right)=\left(\begin{array}{c}
\boldsymbol{\eta}\\
\boldsymbol{\nu}\\
\overline{\boldsymbol{\mu}}
\end{array}\right)-\left(\begin{array}{ccc}
\mathbf{C}^{\eta\eta} & \mathbf{C}^{\eta\nu} & \mathbf{C}^{\times}\\
\mathbf{C}^{\nu\eta} & \mathbf{0} & \mathbf{0}\\
\mathbf{C}^{\times T} & \mathbf{0} & \overline{\mathbf{C}}
\end{array}\right)\left(\begin{array}{c}
0\\
\boldsymbol{\gamma}\\
0
\end{array}\right)
\end{equation}
\begin{equation}
=\left(\begin{array}{c}
\eta_{a}(t)-\sum_{a'}\int_{0}^{\overline{t}}dt'\,C_{aa'}^{\eta\nu}(t,t')\gamma_{a'}(t')\\
\nu_{a}(t)\\
\overline{\mu}_{a}(\tau)
\end{array}\right)\,,
\end{equation}
where $\textnormal{C}_{aa'}^{\eta\nu}=2i\Theta K_{aa\'}^{\textnormal{Im}}$
is the $\eta_{a}-\nu_{a\'}$ correlation function of Eq. \eqref{eq:etanu}.
Hence, when written explicitly, the new $\eta_{a}$ noises are 
\begin{align}
\eta_{a}\'(t)=\eta_{a}(t)-2i\hbar\int_{t_{0}}^{\overline{t}}dt_{1}\Theta(t-t_{1})\sum_{a\'}K_{aa\'}^{\textnormal{Im}}(t-t_{1})\,\text{tr}\left[\mathcal{V}^{a\'}G_{CC}^{+-}(t_{1},t_{1}\vert t_{1})\right],\label{eq:etaprime}
\end{align}
and all the other noises are left unchanged $\nu_{a}\'=\nu_{a}$,
$\overline{\mu}_{a}\'=\overline{\mu}_{a}$.

For the integral over $\mathcal{M}\'$ with respect to the primed
noises to be analytically equivalent to the integral over $\mathcal{M}$
with respect to the original noises, the Jacobian of the transformation
$J$ must be equal to unity. It is convenient to split the Jacobian
matrix into block matrices which correspond to derivatives of the
different noises within the primed and original sets. Using obvious
notation, this is 
\begin{equation}
\mathbf{J}=\begin{vmatrix}\mathbf{J}^{\eta\'\eta} & \mathbf{J}^{\eta\'\nu} & \mathbf{J}^{\eta\'\overline{\mu}}\\
\mathbf{J}^{\nu\'\eta} & \mathbf{J}^{\nu\'\nu} & \mathbf{J}^{\nu\'\overline{\mu}}\\
\mathbf{J}^{\overline{\mu}\'\eta} & \mathbf{J}^{\overline{\mu}\'\nu} & \mathbf{J}^{\overline{\mu}\'\overline{\mu}}
\end{vmatrix}=\begin{vmatrix}\mathbf{J}^{\eta\'\eta} & \mathbf{J}^{\eta\'\nu} & \mathbf{J}^{\eta\'\overline{\mu}}\\
\mathbf{0} & \boldsymbol{\textnormal{I}} & \mathbf{0}\\
\mathbf{0} & \mathbf{0} & \boldsymbol{\textnormal{I}}
\end{vmatrix}\label{eq:J}
\end{equation}
where the second equality is obtained since the $\eta$ noises are
the only set of noises altered by the transformation; so, $\mathbf{J}^{\nu\'\nu}$
and $\mathbf{J}^{\overline{\mu}\'\overline{\mu}}$ are the identity
matrices, while the other blocks in the second and third rows must
be zero. By considering the elements of the $\eta\'\eta$ block matrix,
$J_{ab}^{\eta\'\eta}=\lvert\delta\eta_{a}\'/\delta\eta_{b}(t\')\rvert$,
using Eq. \eqref{eq:etaprime}, and discretising times,
\begin{align}
\frac{\delta\eta_{a}\'(t)}{\delta\eta_{b}(t\')}=\delta_{ab}\delta_{tt'}-2i\hbar\int_{t_{0}}^{\overline{t}}dt_{1}\Theta(t-t_{1})\sum_{a\'}K_{aa\'}^{\textnormal{Im}}(t-t_{1})\,\text{tr}\left[\mathcal{V}^{a\'}\frac{\delta G_{CC}^{+-}(t_{1},t_{1}\vert t_{1})}{\delta\eta_{b}(t\')}\right],\label{eq:Jeta}
\end{align}
it is clear that the Heaviside function bounds the integral from from
above so that $t_{1}<\overline{t}$, while the kernel $K_{aa\'}^{\textnormal{Im}}$
is a known correlation function which is independent of any individual
realization of any specific noise in the set, so the derivative is
applied to the three-time NEGF instead. The NEGF must satisfy causality:
it can only depend on noises from the past, which bounds the integral
from below, $t_{1}>t\'$, meaning that the integral corresponds to
an upper-triangular matrix with zeros on the diagonal, so that its
contribution to the determinant is zero. Therefore the second term
does not contribute to the Jacobian, leaving $J_{ab}^{\eta\'\eta}=\delta_{ab}\delta_{tt'}$
which means that $\mathbf{J}^{\eta\'\eta}$ is the identity matrix.
The other two blocks, $\mathbf{J}^{\eta\'\nu}$ and $\mathbf{J}^{\eta\'\overline{\mu}}$,
are irrelevant since the total Jacobian matrix Eq. \eqref{eq:J} is
an upper triangular matrix with identities along the diagonal and
hence has determinant equal to one. Finally, the simple change of
variables $\eta_{a}\rightarrow\eta_{a}\'$ completes the transformation
with the consequence that the regular stochastic average over realizations
of the noises with the transformed $\{\eta_{a}\'\}$ must be equivalent
to the stochastic average over realizations of the original noises,
in the limit of sampling over all possible realizations. Therefore
Eq. \eqref{eq:Piaverage} becomes 
\begin{align}
\langle\Pi(t)\rangle_{\xi\overline{\xi}}=\langle\Pi_{0}\rangle_{\xi\'\overline{\xi}\'}=\mathbb{N}^{-1},
\end{align}
and so the product of $\mathbb{N}$ and $\langle\Pi_{0}\rangle_{\xi\'\overline{\xi}\'}$
in the first term in Eq. \eqref{eq:lead-L-population} is equal to
one, and the infinite term in the population in Eq. \eqref{eq:lead-L-population}
becomes 
\begin{align}
-i\hbar\mathbb{N}\Pi(\overline{t})\text{tr}\left[g_{LL}^{0<}(\overline{t},\overline{t})\right]=-i\hbar\text{tr}\left[g_{LL}^{0<}(\overline{t},\overline{t})\right]\,,
\end{align}
which is the total number of electrons in the $L$-th lead. Since
$g_{LL}^{0<}(\overline{t},\overline{t})$ does not depend on time
$\overline{t}$, this term is constant and does not contribute to
the current. Concluding, the infinite term in the number operator
Eq. \eqref{eq:lead-L-population} vanishes in the current through
the $L^{th}$ lead after performing the average. This gives the final
form Eq. \eqref{eq:current}.

\section{Self-Energies in the WBA}

\label{A:selfenergies} As well as the regular embedding self-energy,
three additional self energies have been introduced: $\overline{\Sigma}_{CC}$
and $\overline{\Lambda}_{CC}$ in Eq. \eqref{eq:Pibar} for $\overline{\Pi}(\overline{\tau})$,
and $\Upsilon_{CLC}$ in the expression for the current, Eq. \eqref{eq:current}.
Note that the regular embedding self-energy is still required to calculate
the components of the phonon-free NEGF, see, e.g., Ref. \cite{PhysRevB.91.125433}.
The self-energies are reproduced here for convenience (we here set
$\hbar=1$): 
\begin{gather}
\Sigma_{CC}(z_{1},z_{2})=\sum_{L}h_{CL}g_{LL}^{0}(z_{1},z_{2})h_{LC}\\
\overline{\Sigma}_{CC}(\tau_{1},\tau_{2}\vert\overline{\tau})=\sum_{L}h_{CL}\overline{g}_{LL}^{0}(\tau_{1},\tau_{2}\vert\overline{\tau})h_{LC}\\
\overline{\Lambda}_{CC}(\tau_{1},\tau_{2}\vert\overline{\tau})=\sum_{L}h_{CL}\overline{g}_{LL}^{0>}(\tau_{1},0\vert\overline{\tau})h_{LL}^{M}\overline{g}_{LL}^{0<}(0,\tau_{2}\vert\overline{\tau})h_{LC}\\
\Upsilon_{CLC}^{\gamma\gamma\'}(z_{1},z_{2})=h_{CL}g_{LL}^{0\gamma}(z_{1},\overline{t}^{-}\vert\overline{t})g_{LL}^{0\gamma\'}(\overline{t}^{+},z_{2}\vert\overline{t})h_{LC}.
\end{gather}

Using the equation of motion for the isolated lead Green's function
(the index $L$ of the lead is omitted), 
\begin{align}
(i\hbar\partial-h_{LL})g_{LL}^{0}=\delta,
\end{align}
its different components are: 
\begin{gather}
g_{ij}^{>}(t_{1},t_{2})=-\frac{i}{\hbar}\delta_{ij}e^{-i\phi_{i}(t_{1},t_{2})}\left[1-f(\epsilon_{i}^{M})\right]\label{eq:g>}\\
g_{ij}^{<}(t_{1},t_{2})=\frac{i}{\hbar}\delta_{ij}e^{-i\phi_{i}(t_{1},t_{2})}f(\epsilon_{i}^{M})\label{eq:g<}\\
g_{ij}^{r}(t_{1},t_{2})=-\frac{i}{\hbar}\delta_{ij}\Theta(t_{1}-t_{2})e^{-i\phi_{i}(t_{1},t_{2})}\label{eq:gr}\\
g_{ij}^{a}(t_{1},t_{2})=\delta_{ij}\Theta(t_{2}-t_{1})e^{-i\phi_{i}(t_{1},t_{2})}\label{eq:ga}\\
g_{ij}^{\leftGF}(\tau,t)=-\frac{i}{\hbar}\delta_{ij}e^{-\epsilon_{i}^{M}\tau}e^{i\phi_{i}(t,t_{0})}\left[1-f(\epsilon_{i}^{M})\right]\label{eq:gleft}\\
g_{ij}^{\rightGF}(t,\tau)=\frac{i}{\hbar}\delta_{ij}e^{\epsilon_{i}^{M}\tau}e^{-i\phi_{i}(t,t_{0})}f(\epsilon_{i}^{M})\label{eq:gright}\\
g_{ij}^{M}(\tau_{1},\tau_{2})=-\frac{1}{\hbar}\delta_{ij}e^{-\epsilon_{i}^{M}(\tau_{1}-\tau_{2})}\left\{ \Theta(\tau_{1}-\tau_{2})\left[1-f(\epsilon_{i}^{M})\right]-\Theta(\tau_{2}-\tau_{2})f(\epsilon_{i}^{M})\right\} .\label{eq:gM}
\end{gather}
Here, 
\begin{equation}
\phi_{i}(t,t\')=\int_{t\'}^{t}dt_{1}\left[\epsilon_{i}+V_{L}(t_{1})\right]=\epsilon_{i}(t-t\')+\psi_{L}(t,t\')
\end{equation}
for $i\in L$, $f(\omega)=\left(1+e^{\beta\omega}\right)^{-1}$ is
the Fermi function, and $\epsilon_{i}^{M}=\epsilon_{i}-\mu$ as before,
and we also note the useful relationship $[1-f(\omega)]=e^{\beta\omega}f(\omega)$.
There is also the isolated Green's function on the vertical subbranch
up to the preparation time, defined in Eq. \eqref{eq:gbar},
\begin{align}
\overline{g}_{ij}(\tau_{1},\tau_{2}\vert\overline{\tau})=-\frac{1}{\hbar}\delta_{ij}e^{-\epsilon_{i}^{M}(\tau_{1}-\tau_{2})}\left\{ \Theta(\tau_{1}-\tau_{2})\left[1-\overline{f}(\epsilon_{i}^{M}\vert\overline{\tau})\right]-\Theta(\tau_{2}-\tau_{1})\overline{f}(\epsilon_{i}^{M}\vert\overline{\tau})\right\}  & .
\end{align}

All the necessary components of the self-energies can now be obtained.
For the retarded and advanced components which involve the Heaviside
function, the part that depends exclusively on a time difference must
be Fourier transformed to properly account for the Heaviside function.
This gives for the two components:

\[
\Sigma_{nm}^{r}(t_{1},t_{2})=\sum_{L}e^{-i\psi_{L}(t_{1},t_{2})}\int\frac{d\omega}{2\pi}e^{-i\omega\left(t_{1}-t_{2}\right)}\left[\Phi_{nm}(\omega)-\frac{i}{2}\Gamma_{nm}(\omega)\right]\,,
\]
\[
\Sigma_{nm}^{a}(t_{1},t_{2})=\sum_{L}e^{-i\psi_{L}(t_{1},t_{2})}\int\frac{d\omega}{2\pi}e^{-i\omega\left(t_{1}-t_{2}\right)}\left[\Phi_{nm}(\omega)+\frac{i}{2}\Gamma_{nm}(\omega)\right]\,,
\]
where we introduced the level width matrix in terms of the elements
$T_{ni}$ of the transmission matrix of the lead and central regions:
\begin{gather}
\Gamma_{nm}(\omega)=\sum_{L}\Gamma_{nm}^{L}(\omega)=2\pi\sum_{L,i\in L}T_{ni}T_{im}\delta(\epsilon_{i}-\omega)\,,
\end{gather}
and 
\[
\Phi_{nm}(\omega)=\fint\frac{d\omega'}{2\pi}\frac{\Gamma_{nm}(\omega')}{\omega-\omega'}
\]
is basically the Gilbert transform of the level width matrix.

Other components of the self energy are obtained directly by introducing
integration over the energies $\omega$ by means of the delta function
$\delta(\omega-\epsilon_{i})$. This enables one to essentially replace
the sum over the lead's states $i$ with the $\omega$ integration
of the level width matrix: : 
\begin{gather}
\Sigma_{CC}^{>}(t_{1},t_{2})=-ie^{-i\mu(t_{1}-t_{2})}\sum_{L}e^{-i\psi_{L}(t_{1},t_{2})}\int\frac{d\omega}{2\pi}e^{-i\omega(t_{1}-t_{2})}\Gamma_{CC}^{L}(\omega+\mu)\left[1-f(\omega)\right]\label{eq:Sigma>}\\
\Sigma_{CC}^{<}(t_{1},t_{2})=ie^{-i\mu(t_{1}-t_{2})}\sum_{L}e^{-i\psi_{L}(t_{1},t_{2})}\int\frac{d\omega}{2\pi}e^{-i\omega(t_{1}-t_{2})}\Gamma_{CC}^{L}(\omega+\mu)f(\omega)\\
\Sigma_{CC}^{\leftGF}(\tau,t)=-ie^{i\mu(t-t_{0})}\sum_{L}e^{i\psi_{L}(t,t_{0})}\int\frac{d\omega}{2\pi}e^{i\omega(t-t_{0})}e^{-\omega\tau}\Gamma_{CC}^{L}(\omega+\mu)\left[1-f(\omega)\right]\\
\Sigma_{CC}^{\rightGF}(t,\tau)=ie^{-i\mu(t-t_{0})}\sum_{L}e^{-i\psi_{L}(t,t_{0})}\int\frac{d\omega}{2\pi}e^{-i\omega(t-t_{0})}e^{\omega\tau}\Gamma_{CC}^{L}(\omega+\mu)f(\omega)\\
\Sigma_{CC}^{M>}(\tau_{1},\tau_{2})=-i\int\frac{d\omega}{2\pi}e^{-\omega(\tau_{1}-\tau_{2})}\Gamma_{CC}(\omega+\mu)\left[1-f(\omega)\right]\\
\Sigma_{CC}^{M<}(\tau_{1},\tau_{2})=i\int\frac{d\omega}{2\pi}e^{-\omega(\tau_{1}-\tau_{2})}\Gamma_{CC}(\omega+\mu)f(\omega).
\end{gather}

The same is done for the necessary components of the thermal self-energies
on the subbranch up to the preparation time, $\overline{\Sigma}_{CC}$
and $\overline{\Lambda}_{CC}$, by using $\overline{g}_{LL}$ rather
than $g_{LL}^{0M}$, 
\begin{gather}
\overline{\Sigma}_{CC}^{>}(\tau_{1},0^{+}\vert\overline{\tau})=-\int\frac{d\omega}{2\pi}e^{-\omega\tau_{1}}\Gamma_{CC}(\omega+\mu)\left[1-\overline{f}(\omega\vert\overline{\tau})\right]\label{eq:Sigmabar1}\\
\overline{\Sigma}_{CC}^{<}(0,\tau_{1}\vert\overline{\tau})=\int\frac{d\omega}{2\pi}e^{\omega\tau_{1}}\Gamma_{CC}(\omega+\mu)\overline{f}(\omega\vert\overline{\tau})\label{eq:Sigmabar2}\\
\overline{\Lambda}_{CC}(\tau_{1},\tau_{2}\vert\overline{\tau})=-\int\frac{d\omega}{2\pi}\omega e^{-\omega(\tau_{1}-\tau_{2})}\Gamma_{CC}(\omega+\mu)\overline{f}(\omega\vert\overline{\tau})\left[1-\overline{f}(\omega\vert\overline{\tau})\right].\label{eq:Lambdabar}
\end{gather}

Substituting in the $<$, $>$, $\leftGF$ and $\rightGF$ projections
of the isolated lead Green's functions (Eqs. \eqref{eq:g>}, \eqref{eq:g<},
\eqref{eq:gleft}, and \eqref{eq:gright}, respectively) into $\Upsilon_{CLC}^{\gamma\gamma\'}(z_{2},z_{1})$,
where $\gamma$ and $\gamma\'$ represent these projections, and then
introducing the $\omega$ integration by means of $\delta(\omega-\epsilon_{i})$
as above, the components which appear in Eq. \eqref{eq:trYG} are
obtained:
\begin{gather}
\begin{aligned}\Upsilon_{CLC}^{<>}(t_{2},t_{1}) & =e^{-i\mu(t_{2}-t_{1})}e^{-i\psi_{L}(t_{2},t_{1})}\int\frac{d\omega}{2\pi}e^{-i\omega(t_{2}-t_{1})}\Gamma_{CC}^{L}(\omega+\mu)f(\omega)\left[1-f(\omega)\right]\\
 & =\Upsilon_{CLC}^{><}(t_{2},t_{1})
\end{aligned}
\\
\Upsilon_{CLC}^{<<}(t_{2},t_{1})=-e^{-i\mu(t_{2}-t_{1})}e^{-i\psi_{L}(t_{2},t_{1})}\int\frac{d\omega}{2\pi}e^{-i\omega(t_{2}-t_{1})}\Gamma_{CC}^{L}(\omega+\mu)f(\omega)^{2}\\
\Upsilon_{CLC}^{>>}(t_{2},t_{1})=-e^{-i\mu(t_{2}-t_{1})}e^{-i\psi_{L}(t_{2},t_{1})}\int\frac{d\omega}{2\pi}e^{-i\omega(t_{2}-t_{1})}\Gamma_{CC}^{L}(\omega+\mu)\left[1-f(\omega)\right]^{2}\\
\Upsilon_{CLC}^{<\rightGF}(t,\tau)=-e^{-i\mu(t-t_{0})}e^{-i\psi_{L}(t,t_{0})}\int\frac{d\omega}{2\pi}e^{-i\omega(t-t_{0})}e^{\omega\tau}\Gamma_{CC}^{L}(\omega+\mu)f(\omega)^{2}\\
\Upsilon_{CLC}^{>\rightGF}(t,\tau)=e^{-i\mu(t-t_{0})}e^{-i\psi_{L}(t,t_{0})}\int\frac{d\omega}{2\pi}e^{-i\omega(t-t_{0})}e^{\omega\tau}\Gamma_{CC}^{L}(\omega+\mu)f(\omega)\left[1-f(\omega)\right]\\
\Upsilon_{CLC}^{\leftGF<}(\tau,t)=e^{i\mu(t-t_{0})}e^{i\psi_{L}(t,t_{0})}\int\frac{d\omega}{2\pi}e^{i\omega(t-t_{0})}e^{-\omega\tau}\Gamma_{CC}^{L}(\omega+\mu)f(\omega)\left[1-f(\omega)\right]\\
\Upsilon_{CLC}^{\leftGF>}(\tau,t)=-e^{i\mu(t-t_{0})}e^{i\psi_{L}(t,t_{0})}\int\frac{d\omega}{2\pi}e^{i\omega(t-t_{0})}e^{-\omega\tau}\Gamma_{CC}^{L}(\omega+\mu)\left[1-f(\omega)\right]^{2}\\
\Upsilon_{CLC}^{\leftGF\,\rightGF}(\tau_{2},\tau_{1})=\int\frac{d\omega}{2\pi}e^{-\omega(\tau_{2}-\tau_{1})}\Gamma_{CC}^{L}(\omega+\mu)f(\omega)\left[1-f(\omega)\right].
\end{gather}

\bibliographystyle{unsrt}

\end{document}